\documentclass{aastex63}
\usepackage{amssymb}

\usepackage{color}
\usepackage{rotating}
\usepackage{comment}

\newcommand{\chandra}{{CHANDRA}}
\newcommand{\rxte}{{\it RXTE}}

\newcommand{\suzaku}{{\it Suzaku}}
\newcommand{\xmm}{{\it XMM-Newton}}

\newcommand{\ec}{$\eta$~Car}

\newcommand{\swift}{{\it Swift}}
\newcommand{\nicer}{{NICER}}

\newcommand{\ms}{$M_{\odot}$}

\newcommand{\rs}{$R_{\odot}$}

\newcommand{\kms}{km~s$^{-1}$}
\newcommand{\fluxcgs}{ergs~s$^{-1}$~cm$^{-2}$}
\newcommand{\lumcgs}{ergs~s$^{-1}$}

\received{MONTH DAY, YEAR}
\revised{MONTH DAY, YEAR9}
\accepted{\today}
\submitjournal{ApJ}

\shorttitle{\nicer\ X-ray Variations of \ec}
\shortauthors{Espinoza et al.}

\begin{document}

\title{\nicer\ X-ray Observations of Eta Carinae During its Most Recent Periastron Passage}

\correspondingauthor{David Espinoza-Galeas}
\email{84espinozagaleas@cua.edu}

\author{David Espinoza-Galeas}
\affiliation{Institute for Astrophysics and Computational Sciences, The Catholic University of America, 620 Michigan Ave., N.E. Washington, DC 20064}
\affiliation{Departamento de Astronomía y Astrof\'{i}sica, Facultad de Ciencias Espaciales, Universidad Nacional Autónoma de Honduras, Bulevar Suyapa, Tegucigalpa, M.D.C, Honduras, Centroam\'{e}rica }

\author{M.~F.~Corcoran}
\affiliation{Institute for Astrophysics and Computational Sciences, The Catholic University of America, 620 Michigan Ave., N.E. Washington, DC 20064}
\affiliation{CRESST and X-ray Astrophysics Laboratory, NASA/Goddard Space Flight Center, Greenbelt, MD 20771}

\author{K.~Hamaguchi}
\affiliation{CRESST and X-ray Astrophysics Laboratory, NASA/Goddard Space Flight Center, Greenbelt, MD 20771}
\affiliation{Department of Physics, University of Maryland, Baltimore County, 1000 Hilltop Circle, Baltimore, MD 21250, USA.}

\author{C.~M.~P.~Russell}
%\affiliation{The Catholic University of America, 620 Michigan Ave., N.E. Washington, DC 20064}
\affiliation{Institute for Astrophysics and Computational Sciences, The Catholic University of America, 620 Michigan Ave., N.E. Washington, DC 20064}
\affiliation{Department of Physics and Astronomy, University of Delaware, Newark, DE 19716, USA}

\author{T.~R.~Gull}
\affiliation{Exoplanets \&\ Stellar Astrophysics Laboratory, NASA/Goddard Space Flight Center, Greenbelt, MD 20771, USA}
\affiliation{Space Telescope Science Institute, 3700 San Martin Drive. Baltimore, MD 21218, USA}

\author{A.~F.~J.~Moffat}
\affiliation{D\'epartement de physique and Centre de Recherche en Astrophysique du Qu\'ebec (CRAQ), Universit\'e de Montr\'eal, C.P. 6128, Succ.~Centre-Ville, Montr\'eal, Qu\'ebec, H3C 3J7, Canada}

\author{N.~D.~Richardson}
\affiliation{Embry Department of Physics and Astronomy, Embry-Riddle Aeronautical University, 3700 Willow Creek Rd, Prescott, AZ, 86301, US}

\author{G. Weigelt}
\affiliation{Max-Planck-Institut f\"ur Radioastronomie, Auf dem H\"ugel 69, 53121 Bonn, Germany}

\author{D.~John~Hillier}
\affiliation{Department of Physics and Astronomy \& Pittsburgh Particle Physics, Astrophysics and Cosmology Center (PITT PACC), \\ \hspace{1cm}  University of Pittsburgh, 3941 O'Hara Street,  Pittsburgh, PA 15260, USA }

\author{Augusto Damineli}
\affiliation{Universidade de S\~{a}o Paulo, IAG, Rua do Mat\~{a}o 1226, Cidade Universit\'{a}ria S\~{a}o Paulo-SP, 05508-090, Brasil}

\author{Ian~R.~Stevens}
\affiliation{School of Physics \& Astronomy, University of Birmingham, Birmingham B15 2TT, UK}

\author{Thomas~Madura}
\affiliation{Department of Physics \& Astronomy, San Jose State University, One Washington Square, San Jose, CA 95192, USA}

\author{K.~Gendreau}
\affiliation{X-ray Astrophysics Laboratory, NASA/Goddard Space Flight Center, Greenbelt, MD 20771}

\author{Z.~Arzoumanian}
\affiliation{X-ray Astrophysics Laboratory, NASA/Goddard Space Flight Center, Greenbelt, MD 20771}

\author{Felipe Navarete}
\affiliation{SOAR Telescope/NSF's NOIRLab\\ Avda Juan Cisternas 1500, 1700000, La Serena, Chile}

\begin{abstract}

We report high-precision X-ray monitoring observations in the 0.4--10~keV band of the luminous, long-period colliding-wind binary Eta~Carinae up to and through its most recent X-ray minimum/periastron passage in February 2020.
Eta~Carinae reached its observed maximum X-ray flux on 7~January~2020, at a flux level of $3.30 \times 10^{-10}$~\fluxcgs, followed by a rapid plunge to its observed minimum flux, $0.03 \times 10^{-10}$~\fluxcgs\ near 17~February~2020.
The \nicer\ observations show an X-ray recovery from minimum of only $\sim$16 days, the shortest X-ray minimum observed so far. We provide new constraints of the ``deep'' and ``shallow'' minimum intervals.
Variations in the characteristic X-ray temperature of the hottest observed X-ray emission indicate that the apex of the wind-wind ``bow shock'' enters the companion's wind acceleration zone about 81 days before the start of the X-ray minimum.
There is a step-like increase in column density just before the X-ray minimum, probably associated with the presence of dense clumps near the shock apex.
During recovery and after, the column density shows a smooth decline, which agrees with previous $N_{H}$ measurements made by \swift\ at the same orbital phase, indicating that changes in mass-loss rate are only a few percent over the two cycles.
Finally, we use the variations in the X-ray flux of the outer ejecta seen by \nicer\ to derive a kinetic X-ray luminosity of the ejecta of $\sim 10^{41}$ \lumcgs near the time of the ``Great Eruption''.

\end{abstract}


\section{Introduction} \label{sec:intro}

At  a distance of $\sim$2.3 kpc \citep{2006ApJ...644.1151S}, Eta~Carinae \cite[\ec, ][]{Davidson:1971hb, 1997ARAA..35....1D, Humphreys:2012pd}  is the nearest star system with a mass $\gtrsim$ 100 \ms\ \citep{2001ApJ...553..837H, Madura:2012qf, Corcoran:2012ve}.
The \ec\  system is a high-mass non-compact binary system containing an extremely unstable ``Luminous Blue Variable'' (LBV), \ec-A.
In the mid-nineteenth century, this star experienced (and survived) the most energetic stellar mass ejection event ever observed to date, where a mass of $\gtrsim$ 45 \ms \citep{Morris:2017sf} was ejected.
This material created the bipolar ``Homunculus Nebula''  \citep{1950ApJ...111..408G}, which today surrounds the star and which is expanding outward at a velocity of 650~\kms \citep{1997ARAA..35....1D}.
Observations of the He~I 1.083$\mu$m emission line over a period of 50 years reported by \cite{1996ApJ...460L..49D} established, for the first time, a strict periodicity associated with the stellar spectrum, suggesting that the star is a massive binary system with an orbital period of 5.52 years  \citep{1997NewA....2..107D}.  
The discovery of periodic X-ray variability \citep{Corcoran:1995fk, 1999ApJ...524..983I, 2005AJ....129.2018C} was critical to identifying the system as a highly eccentric ($e\approx0.9$) ``colliding-wind'' binary \citep{1998ApJ...494..381C}.
The X-ray emission observed from the \ec\ system is dominated by hot shocked gas produced when the dense, slow  \citep[$V_{\infty}\approx$ 420~\kms, ][]{Groh:2012fk} wind of the LBV primary \ec-A collides with the wind of its companion star, \ec-B.
Because the companion star has never been directly detected, X-ray observations remain critical to constraining its stellar wind properties and monitoring changes in mass-loss rate from either star.
The temperatures inferred from the X-ray spectrum and the X-ray brightness of the gas shocked in the wind collision suggest that the unseen companion star has a fast \citep[$V_{\infty}\approx$ 3000~\kms, ][]{2002AA...383..636P} lower-density wind, probably some type of hot massive star \citep{2005ApJ...624..973V, Mehner:2010rm}.
Table~\ref{tab:ec_par} summarizes the stellar and orbital parameters of the \ec\ system.

\begin{deluxetable}{rrr}
\label{tab:sysparams}
\tabletypesize{\scriptsize}
\tablecolumns{3}
\tablewidth{0pc}
\tablecaption{\label{tab:ec_par}\ec\ Adopted parameters}
\tablehead{
\colhead{Parameter} &	\colhead{Value} &	\colhead{Reference}}
\startdata
Distance   (pc)         & 2300$\pm$200     & \cite{2006ApJ...644.1151S} \\% \cite{1997ARAA..35....1D}  \\
Period  (X-ray, days)            & 2023.40 $\pm$ 0.71 days               &\cite{2017ApJ...838...45C}\\
Eccentricity       & 0.9                                   & \cite{2001ApJ...547.1034C} \\
$T_{o}$ (Periastron Passage, MJD) &  56873.90  & \cite{2016ApJ...819..131T} \\
$T_{x}$ (Deep X-ray Minimum Start, MJD) &  50799.42 & \cite{2017ApJ...838...45C} \\
Total Luminosity  ($10^{6} L_{\odot}$) & 5      &\cite{2001ApJ...553..837H}\\
Mass, $\eta_A$ (\ms)              & $>100$                         & \cite{2001ApJ...553..837H}\\
Mass, $\eta_B$, ZAMS (\ms) & 40--50 & \cite{Mehner:2010rm} \\ 
$V_{\infty,A}$ (km~s$^{-1}$)  & 420     & \cite{Groh:2012fk} \\
$V_{\infty,B}$ (km~s$^{-1}$)  & 3000        & \cite{2002AA...383..636P} \\
$\dot{M}_{A}$ ($10^{-5}$~\ms~yr$^{-1}$) & $\approx 85$  &\cite{Madura:2013fj} \\
$\dot{M}_{B}$ ($10^{-5}$~\ms~yr$^{-1}$) & $\approx 1.4$ &  \cite{2009MNRAS.394.1758P} \\
$a$  (AU) &  15.9 &  
\enddata
\end{deluxetable}

%The \ec\ system is the nearest massive stellar system where it is possible to study mass loss and its effects on the evolution of the system. 
X-ray spectra provide important observational constraints on temperatures and densities in the hot shocked gas in the wind-wind collision ``bow shock'', while the absorption suffered by the X-ray emission provides information on the density distribution of the wind of the LBV primary (since the matter along the line-of sight to the X-ray source is dominated by the wind of the LBV). These quantities change with the distance between the stars \citep{Stevens:1992yu} as the stars revolve in their extremely eccentric orbit. 
%From these measurements, densities, velocities and mass-loss rates of the winds from the two stars can be inferred \citep{Stevens:1992yu, usov92, Madura:2013fj}, helping us to build a better idea of the current mass loss in the \ec\ system. 

\begin{figure}[htbp]
  \centering
  \includegraphics[width=\linewidth]{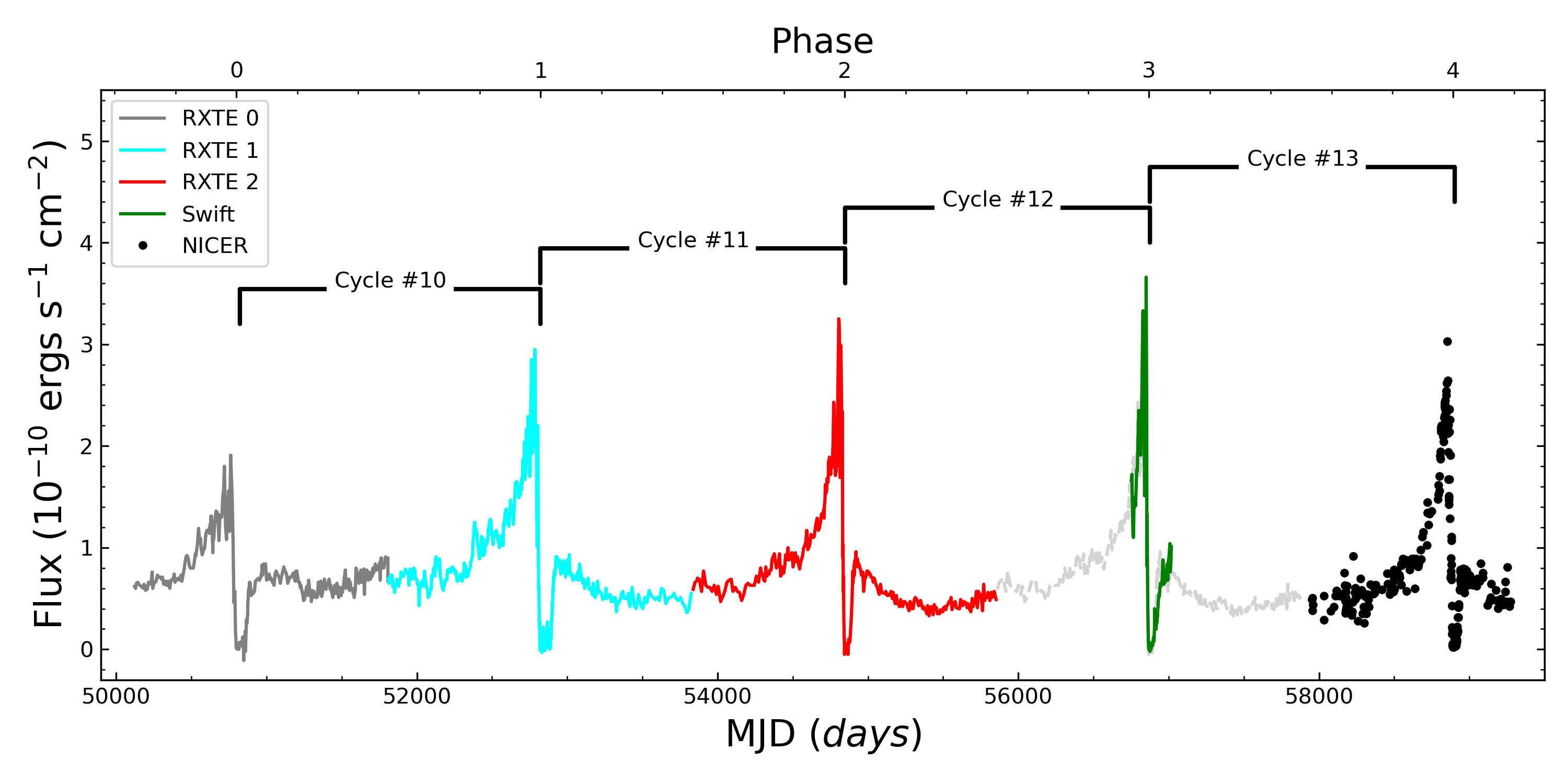}
  \caption{\ec’s X-ray lightcurve between 2.0 - 10.0 keV for \rxte\ and \swift.
    For an easy identification the observations are color code and separated in different sets around the X-ray minima.
    We define $\phi=0$ at the X-ray minimum of the first \rxte\ cycle of observations using epoch and period calculated in \cite{2017ApJ...838...45C}. 
    \cite{2017ApJ...838...45C} used the full set of data to find an X-ray periodicity of 2023.40 $\pm$ 0.71 days which is very close to the periodicity of 2022.7 days found by \cite{2016ApJ...819..131T} for \ec.
The brackets show the corresponding cycle numbers using He I and II minima events in \cite{2008MNRAS.384.1649D} and \cite{2016ApJ...819..131T}.
The light gray dashed line is a repetition of the last \rxte\ cycle.
The plot shows us the variability of the emission in five continuous cycles from 1996 to 2021, including the \nicer\ observations in black.
}
  \label{fig:eta_car_rxte_swift_lc}

\end{figure}

Figure \ref{fig:eta_car_rxte_swift_lc} shows previous observations \citep{1999ApJ...524..983I, 2001ApJ...547.1034C, 2002AA...383..636P, 2005AJ....129.2018C, 2010ApJ...725.1528C, 2017ApJ...838...45C} of the time-variable X-ray emission from the \ec\ system.
%To facilitate the identification of the different set of X-ray observations we separated them in different sets around X-ray minima. 
%The observations around the X-ray minima of 1997, 2003, 2009, and 2015 are in gray, cyan, red and green respectively. 
These observations have helped to establish the temperatures and densities of the shocked gas in the wind-wind collision region, and analyses of these observations using even-more-sophisticated models \citep{2008MNRAS.388L..39O,2011ApJ...726..105P,Madura:2013fj} determined the geometry of the shocked winds, temperatures at the colliding-wind region (CWR) and mass-loss rate of the winds from the stars. 
Even though published 3-D models of the X-ray spectral variability provide a good overall description of the CWR \citep{2011ApJ...726..105P,Russell:2016qy} they do not explain some important details of \ec's X-ray behavior, such as the ``flaring'' episodes seen prior to the start of the X-ray minimum \citep{1999ApJ...524..983I, 2009ApJ...707..693M}, variation in the recovery from the X-ray minimum \citep{2010ApJ...725.1528C}, or the ``Deep''/``Shallow'' minimum  transition \citep{Hamaguchi:2014lr}.   

In this paper we report the first monitoring of the variable X-ray spectrum of \ec\ with the Neutron Star Interior Composition Explorer (\nicer).
Here we discuss the observed X-ray spectrum variations from our \nicer\ X-ray observing campaign, and compare them to X-ray variations seen during previous orbital cycles.
Throughout the paper, phases are calculated using the X-ray ephemeris  \citep{2017ApJ...838...45C} where $E$ is the cycle count. 

\begin{equation}
  \mbox{MJD (X-ray Minimum)} = 50799.4 +2023.4E 
\label{eq:ephem}
\end{equation}

The X-ray epoch $\phi=0$ in Figure \ref{fig:eta_car_rxte_swift_lc} refers to the first \rxte\ deep minimum.
Previous works like \cite{2008MNRAS.384.1649D} and \cite{2016ApJ...819..131T} have used spectral variations of He emission to  number the cycles of \ec, starting from the event observed in 1948 by \cite{1953ApJ...118..234G}.
The square brackets in Figure \ref{fig:eta_car_rxte_swift_lc} show the corresponding cycle number based on the He minima.
The epoch of periastron using He cycles in the Figure is based on variations of the transient He~II $\lambda 4686$ line, $T_{o} = 56873.9$ found by \cite{2016ApJ...819..131T}.
%Figure \ref{fig:eta_car_rxte_swift_lc} also shows the corresponding X-ray cycle number to He EW minima events.
This He II epoch corresponds to the starting point of cycle 10, which is 4.3 days later than the X-ray minimum for the 1997 periastron passage.

This paper is organized as follows.
We describe the \nicer\ observing program and reduction and analysis of the \nicer\  spectra in Section \ref{sec:obs}, including a discussion of \nicer\ background estimation.
We present the results of the analysis of the net X-ray spectra in Section \ref{sec:spec}.
We discuss the main  results of our spectral analysis in Section \ref{sec:disc}, including a comparison of the \nicer\ spectral properties with similar properties from previous cycles, and in particular compare the flux variations to refine the X-ray period, and examine variations prior to and after the start of the X-ray minimum.
We summarize our results in Section \ref{sec:conc} and discuss areas of future work.

\section{Observations}
\label{sec:obs}

\subsection{Description of the Instrument}

The Neutron Star Interior Composition Explorer  \citep[\nicer,][]{Gendreau:2012fv, 2014SPIE.9144E..20A} is an X-ray astronomy facility attached to the International Space Station (ISS).
\nicer\ is devoted to time-resolved X-ray spectrometry in the 0.2~--~12.0 keV energy band.
\nicer\ was launched on 3~June~2017 aboard a Space X Falcon 9 rocket, and deployed at the ISS on 16~June~2017.
\nicer's X-ray Timing Instrument \citep[XTI,][]{Prigozhin:2016lr} is comprised of a co-aligned array of 52 Focal Plane Modules, each consisting of a matched pair of X-ray ``concentrator'' (XRC) optics with a silicon drift detector (SDD) to record the energy and time-of-arrival of source X-ray photons.
Each XRC optic collects X-rays over a roughly 30 arcmin$^{2}$ region of the sky centered on the target of interest in the 0.2~--~12~keV energy band and concentrates them onto an SDD.
\nicer's primary mission is to obtain X-ray spectrometry with high time and moderate spectral resolution of cosmic X-ray sources, primarily of X-ray binary pulsars.
\nicer's combination of large effective area, restricted field of view and broad bandpass in the thermal X-ray range makes it especially well-suited to observe X-ray variables like long-period colliding-wind binaries (and other sources) in addition to X-ray pulsars.
Table \ref{tab:nicer_facts} lists key technical specifications of \nicer's capabilities.

\nicer\ observed \ec\ starting on 20 July 2017 (as part of \nicer's  ``observatory science'' program), continuing with guest investigator programs in Cycle 1 (Principle Investigator, M.~F.~Corcoran) and Cycle 2 (Principal Investigator, D.~Espinoza-Galeas).  
The whole set of \nicer\ observations are listed in the appendix but we present a summary of the observations in Table \ref{tab:obs_sum}.
%Table \ref{tab:nicer-rates} and
\nicer\ has provided frequent measure of \ec's X-ray spectrum, typically twice per month, with appropriately increased sampling near periastron when the variation in the X-ray spectrum occurs on timescales of days. 
An alternative analysis of some of these spectra was discussed by \cite{2021ApJ...914...47K}.

\begin{deluxetable}{lr}\label{tab:nicer_facts}
\tabletypesize{\scriptsize}
\tablecolumns{2}
\tablewidth{0pc}
\tablecaption{\nicer\ XTI Characteristics}
\tablehead{
%\vspace{-0.55cm}
\colhead{Property} & \colhead{ }
}
\startdata
Effective area & $>$2000 cm$^{2}$ at 1.5 keV\\
               & 600 cm$^{2}$ at 6 keV\\
Energy resolution & 85 eV at 1 keV\\
               & 137 eV at 6 keV\\
%Spatial resolution & 5 arcmin\\ 
Broad Bandpass & 0.2 $<$ E $<$ 12.0 keV\\
Absolute timing precision  & $<$ 300 ns\\
%Moderate spectral Resolution & 6 $<$ E/$\Delta$E $<$ 80 from 0.5 keV to 8 keV\\
Restricted field of view & 30 arcmin$^{2}$\\
Sensitivity & $3\times 10^{-14}$\, ergs\, s$^{-1}$\, cm$^{-2}$\\
& (0.5-10.0 keV, 5$\sigma$ in 10 ksec)\\
\enddata
\end{deluxetable}

\begin{deluxetable}{rrrr}
\tabletypesize{\scriptsize}
\tablecolumns{4}
\tablewidth{0pc}
\tablecaption{\ec\ Observation Summary\label{tab:obs_sum}}
\tablehead{
\colhead{} &	\colhead{\# of Obs.} &	\colhead{First Obs.} & \colhead{Last Obs.}}
\startdata
 \chandra\ & 31       & 2000-11-19 & 2020-01-26\\
 \nicer\ & 249    & 2017-07-21 & 2021-02-21\\
\enddata
\end{deluxetable}

\subsection{Data Reduction and Calibration}

The \nicer\  spectra for all \ec\ observations were extracted from the clean, merged photon events file, using data obtained outside of the South Atlantic Anomaly at sun angles $>40^{\circ}$ to avoid optical stray-light contamination, using the \texttt{NICERDAS}\footnote{see \url{https://heasarc.gsfc.nasa.gov/docs/software/lheasoft/help/nicer.html}} software package distributed with HEASoft software analysis package (version 6.27.2).
We used standard \nicer\ event cleaning criteria\footnote{see \url{https://heasarc.gsfc.nasa.gov/lheasoft/ftools/headas/nimaketime.html} for a description of the standard cleaning criteria} to convert the observed event times and pulse-heights to cleaned events with calibrated energies, using data from all 52 active detectors.
We used calibration data (effective areas and instrument response functions) from the publicly available \nicer\ calibration data version 20200202.

\subsection{\nicer\ Background Estimation}

\nicer\ is subject to a varying charged particle environment in the high-inclination ISS orbit, that traverses regions of trapped charged particles near the South Atlantic Anomaly and the regions near the north and south poles (the ``polar horns'').
Charged particle background is most noticeable at high energies.
At low energies, the \nicer\  background is dominated by optical light contamination at low sun angles and an instrumental ``noise peak'' of excess events at energies $<~0.4$~keV.
Correction for background is important, especially near X-ray minimum.  To minimize background contamination in the \nicer\ spectra, we extracted lightcurves in the $0.4-10$~keV band for all the \nicer\ observations and visually inspected them, defining time regions to exclude short intervals of large rapid increases in count rate produced by the variable charged particle environment.
We then extracted X-ray spectra for each observation from the cleaned and screened event files.  
Residual charged particle background and/or optical light contamination generally affects spectra even after the exclusion of obvious intervals of high background. 

Background estimation is still under development, so we used  two different background estimator tools provided by the \nicer\ Guest Observer Facility\footnote{See \url{https://heasarc.gsfc.nasa.gov/docs/nicer/tools/nicer_bkg_est_tools.html}.}, the \texttt{nibackgen3C50} \citep{2021arXiv210509901R} and the  \texttt{nicer\_bkg\_estimator} ``space-weather'' (SW) background estimators.
Both estimators use \nicer\ observations of ``blank sky'' fields (i.e. fields with no obvious sources in the XRC field of view) to estimate background.
The \texttt{nibackgen3C50} tool compares \nicer\ spectra in various bands in the observed and blank-field observations to construct a background, while the \texttt{nicer\_bkg\_estimator} tool matches sun angle, cut off rigidity and the solar Kp index  between the observation and \nicer\ observations of ``blank sky'' regions to estimate the amount of background due to optical loading and the charged particle environment.
Because background estimates continue to be refined, we used both models to gauge the effects of background contamination, but because the SW background estimator showed more dispersion in the $\chi^{2}_{red}$ value of the fittings, we adopted the \texttt{nibackgen3C50} backgrounds in our analysis.

\begin{figure}[htbp]
  \centering
  \includegraphics[width=\linewidth]{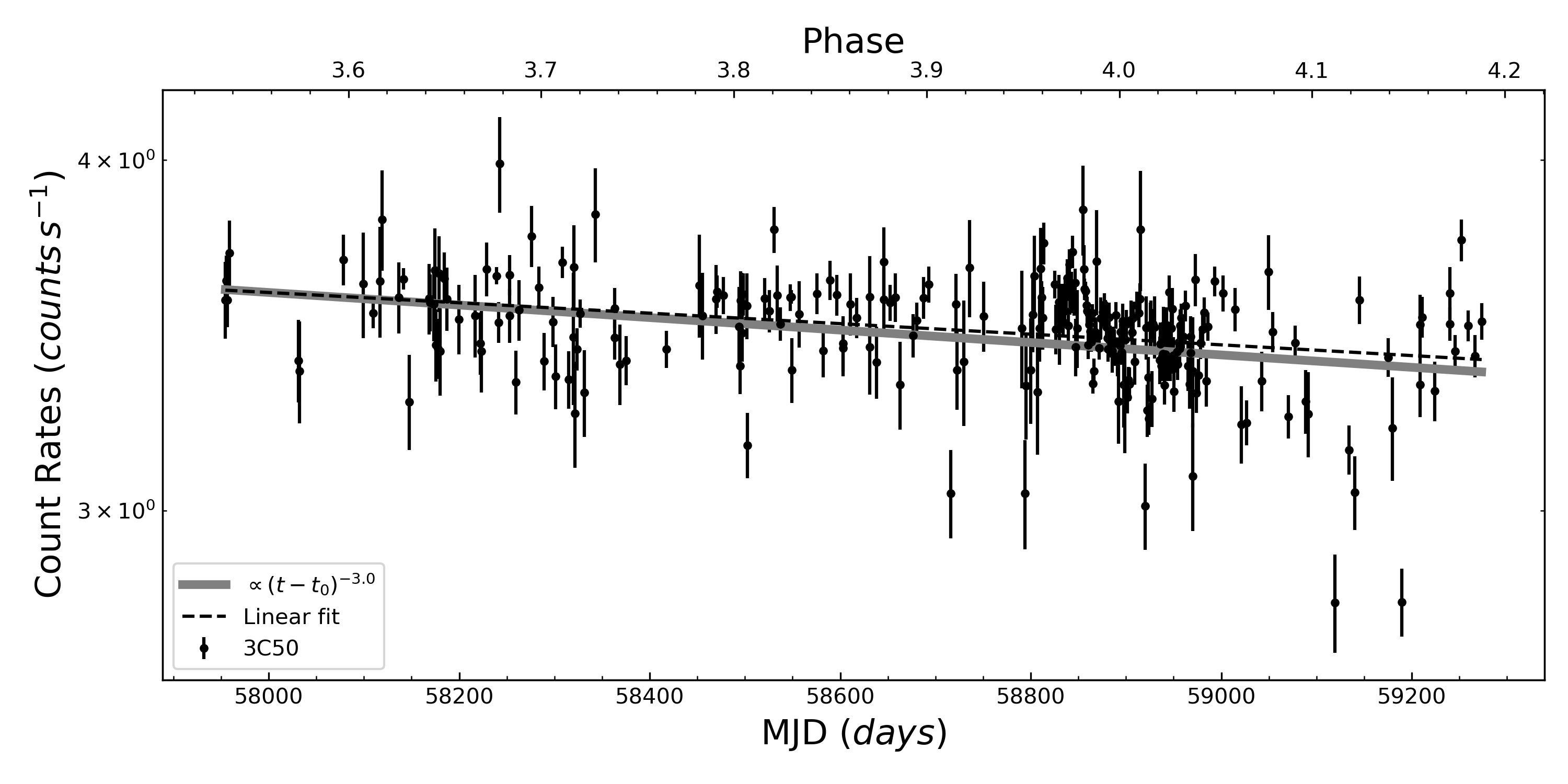}
  \caption{Count rates for \ec\ between 0.5~--~1.0 keV observed by \nicer\ after background correction using the \texttt{nibackgen3C50} model. Emission in this soft band is produced by the shocked gas in the extended outer debris region around \ec, and is expected to be constant over timescales of months-years. The slight decline of \ec's soft X-ray emission (dashed black line) shows good agreement with the $\propto (t-t_{0})^{-3}$ power law (gray line), assuming that the ejecta originated near the time of the Great Eruption in 1843 \citep{Kiminki:2016kq}.}
  \label{fig:eta_car_nicer_0p5_1p0}
\end{figure}

As a check on the background estimation, Figure \ref{fig:eta_car_nicer_0p5_1p0} shows the total count rates for \ec\ between 0.5~--~1.0 keV, along with the net rates using the backgrounds estimated from the \texttt{nibackgen3C50} method.
In this range, the X-ray emission is totally dominated by the emission from the ``Outer Ejecta'' (OE) region surrounding the Homunculus Nebula \citep{1979ApJ...234L..55S, 2004AA...415..595W, 2007ASPC..367..257H}.
The emission from the OE region, which extends out to  $\sim1'$ from \ec,  is entirely within the $1.7'$ radius of the field of view of the \nicer\ SDDs and so present in all  \nicer\ observations of \ec. 
The mean of the total source+background rates in the 0.5~--~1.0~keV is $\mu_{total}= 3.76\pm 0.15$~cts~s$^{-1}$, while the mean of the net count rate using the \texttt{nibackgen3C50} method is $\mu_{net, 3c50}=3.48 \pm 0.14$~cts~s$^{-1}$ and the mean of the net count rate using the space-weather  \texttt{nicer\_bkg\_estimator} method is  $\mu_{net, SW}=3.55\pm 0.20$~cts~s$^{-1}$. In this band the 3c50 method appears to have a slightly smaller standard deviation, but similar to that in the total band.  
Figure \ref{fig:eta_car_nicer_0p5_1p0} shows the net \nicer\ count rates in the 0.5~--~1.0~keV band using the \texttt{nibackgen3C50} background estimate. 
The net \nicer\ count rates in this band seem to show a decline of about 7\% over the $\sim$1000 days of the \nicer\ monitoring, also visible in the gross rates and the net rates corrected using the ``space weather'' model. 
A linear fit to the total (non-background subtracted) count rate data yields a decline of  
$\Delta R/\Delta t = 0.150 \pm 0.028$~cts~ks$^{-1}$~day$^{-1}$.
We suspected this apparent decline is instrumental, as might be expected from increasing condensations on the FPMs, but examination of \nicer\ observations of the supernova remnant 1E~0102.2-7219, an extended soft source, covering the same timespan as the \ec\ observations did not find a comparable decline in the soft-band flux.
This decrease might be due to expansion of the outer ejecta.
\ec's X-ray emission below 2.0 keV comes from gas heated by the shocked ejecta from the 19th century eruption.
The X-ray flux is proportional to the emission measure $F_{X}\propto n^{2}V$, where $n$ is the particle density and $V$ is the emission volume of hot gas. Assuming a constant mass $M$ of gas expanding freely at a constant velocity $\upsilon$ then $F_{X}\propto (t-t_{0})^{-3}$. Figure \ref{fig:eta_car_nicer_0p5_1p0} shows how the $(t-t_{0})^{-3}$ power-law gives good agreement of the observed \ec's soft X-ray emission assuming the gas expansion started near the time of the Great Eruption in 1843. The $(t-t_{0})^{-3}$ decline in X-ray flux is observed in the soft X-ray emission ($0.3-2.0$ keV) of young SN remnants \citep{2005ApJ...632L..99I}. Extrapolating this power law to 100 days after the Great Eruption, the initial X-ray luminosity of the shocked ejecta would have been $L_{x,0}\sim 10^{41}$ \lumcgs. This would be close to the total luminosity at longer wavelengths derived from the observed brightness of the system at that time (L=$10^{7.3}$~L$_{\odot}$, \cite{1997ARAA..35....1D}). If confirmed, this would be the first estimate of the X-ray luminosity of the Great Eruption.

\section{Multi-Band X-ray Variations}
\label{sec:crv}

\subsection{\nicer\ Broad Band X-ray Lightcurves of \ec}

\begin{figure}[htbp]
  \centering
  \includegraphics[width=\linewidth]{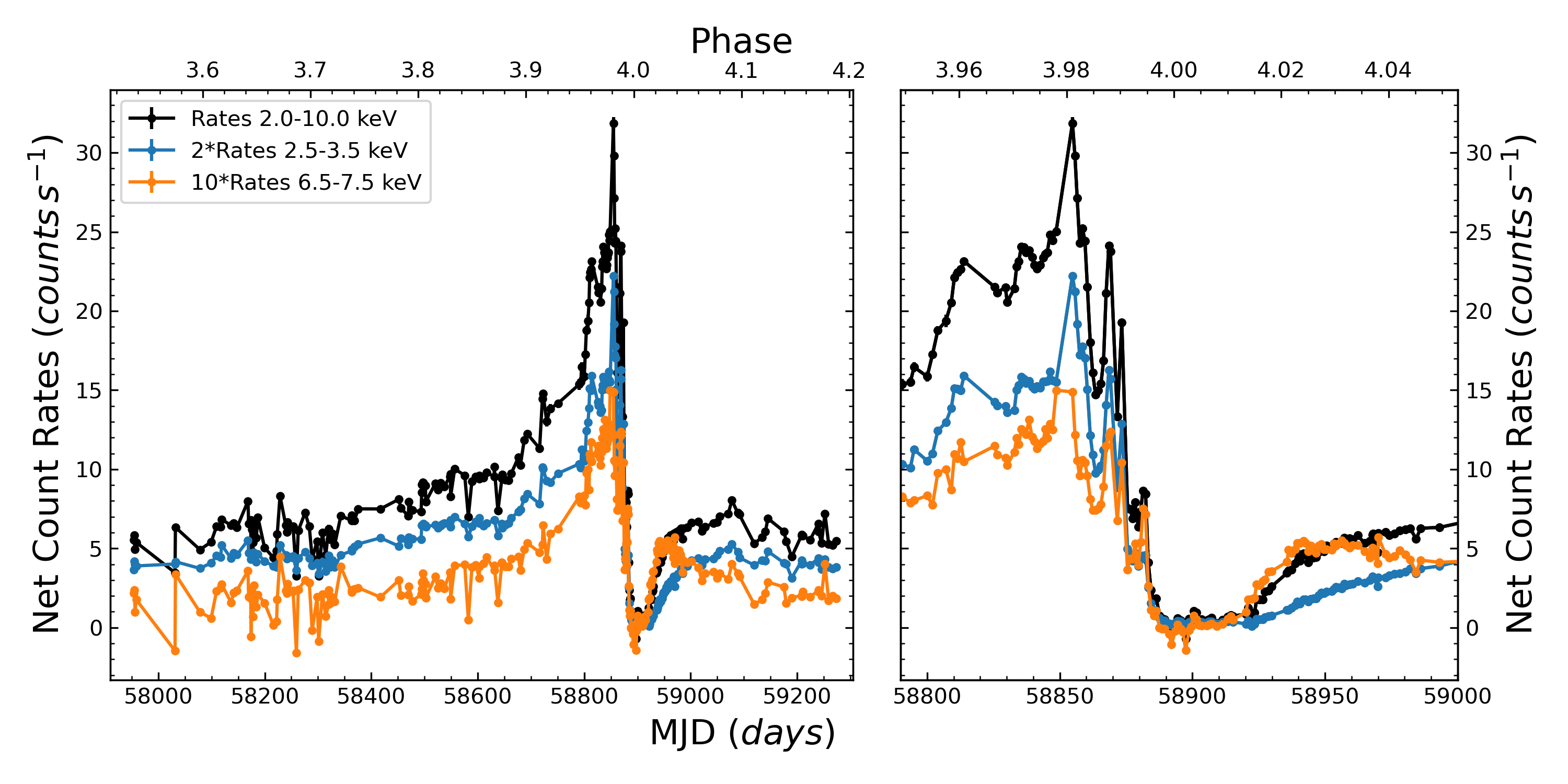}
  \caption{\nicer\ net count rates for \ec\ using the ``3c50'' background model, in three energy bands. The 2.5~--~3.5 keV and 6.5~--~7.5~keV count rates have been multiplied by $2\times$ and $10\times$, respectively, for display purposes.}
  \label{fig:eta_car_nicer_crts}
\end{figure}

Measurements of net count rates in the 2.5~--~3.5 keV, 6.5~--~7.5 keV, and 2.0~--~10.0 keV bands are shown in Figure~\ref{fig:eta_car_nicer_crts}.
Rates in these three bands are dominated by emission from the hot gas in the colliding-wind shock and background contamination is fairly low.
The 2.0~--~10.0 keV band was originally adopted by \cite{1999ApJ...524..983I} in their analysis of the \rxte\ monitoring observations of \ec, and includes nearly all of the observable thermal X-ray emission from the colliding-wind region, while excluding contamination from the OE region which is of lower energy.

The net count rates soon after the start of the \nicer\ observations (on MJD 57954.5, 2017-07-20) show significant artificial scatter from observation to observation because of variations in background which are not accurately accounted for by the background model.
To mitigate this variability, we requested that observations obtained after MJD 58306 (2018-07-07) be restricted to avoid times of low geomagnetic rigidity, i.e. when the ISS was not near the South Atlantic Anomaly or the polar horns.
This restriction significantly reduced variability from charged particle events, which accounts for the reduced scatter after that time.   

The X-ray variability seen by \nicer\ in the 2~--~10 keV band is similar to that seen in other cycles by the \rxte\ Proportional Counter Array and \swift\ X-ray Telescope.
As shown in Figure~\ref{fig:eta_car_nicer_crts}, the count rates in all energy bands show a gradual increase from MJD 57954.5 (2017-07-20) to MJD 58610 (2019-05-07).
After MJD 58610, the count rate increase accelerates, and short-term brightenings \citep[``flares'',][]{2009ApJ...707..693M} occur as the stars approach X-ray minimum near periastron passage. 
Starting at MJD 58855 (2020-01-06, about  $\phi\sim$0.98) a rapid decline in X-ray count rate began, reaching to the ``Deep'' X-ray minimum near MJD 58893 (2020-02-14).
The emission in this band starts to increase by MJD 58912.5  (2020-03-04), just 19.5 days after the start of the Deep minimum.

The 6.5~--~7.5 keV band is dominated by the high energy thermal emission and includes emission in the Fe~K region.
This band is not greatly affected by absorption from the wind of the LBV except at phases very close to the X-ray minimum when column densities rise in excess of $10^{23}$~cm$^{-2}$. 
The flares in the 6.5~--~7.5 keV band are somewhat smaller in amplitude compared to the 2~--~10 keV band, and there is also evidence that the flares begin sooner in the 6.5~--~7.5 keV than at lower energies.
All the flares are present in the 2.5~--~3.5 and 6.5~--~7.5 bands, except for the last flare just before X-ray minimum.
The last flare occurs only in the 6.5~--~7.5 band while the 2.5~--~3.5 count rates stay almost constant.
The plunge to the X-ray minimum in the  6.5~--~7.5 keV band seems to occur almost simultaneously with the drop seen in the lower-energy bands, and X-ray minimum flux is reached at MJD 58854 (2020-01-06) in all energy bands.
The X-ray count rates in the all bands begin to increase by MJD 58912.5 (2020-03-04), but at different rates.
The rate of increase is greatest in the 6.5~--~7.5 keV band, and net rates in this band increase until 58942 (2020-04-03), when it begins to start a slow decline as the stars separate towards apastron.
Of the three bands, the rate of increase is slowest in the low energy 2.5~--~3.5 keV band, which shows  the importance of extended absorption of the low energy X-rays by the intervening, unshocked wind of \ec-A.
This absorption continues for about 200 days after periastron passage.

\subsection{Hardness Variations}

We calculated the hardness ratio $HR = (H-M)/(H+M)$  where $H$, the hard band, is defined as the net count rates between $6.5<E<7.5$~keV, and the medium energy band, $M$, the net count rates between $2.5<E<3.5$~keV.
We chose these bands because both are dominated by the colliding-wind source with (generally) minimal background contamination, and because the medium band is also sensitive to changes in absorption (as shown in Figure~\ref{fig:eta_car_nicer_spectrum}) more so than the hard band.
This hardness ratio (HR) mainly provides a measure of how the absorption to the X-ray source is changing in time, since changes in shock temperature should be modest, because for most of the orbit the winds collide at terminal velocity.

\begin{figure}[htbp]
  \centering
  \includegraphics[width=\linewidth]{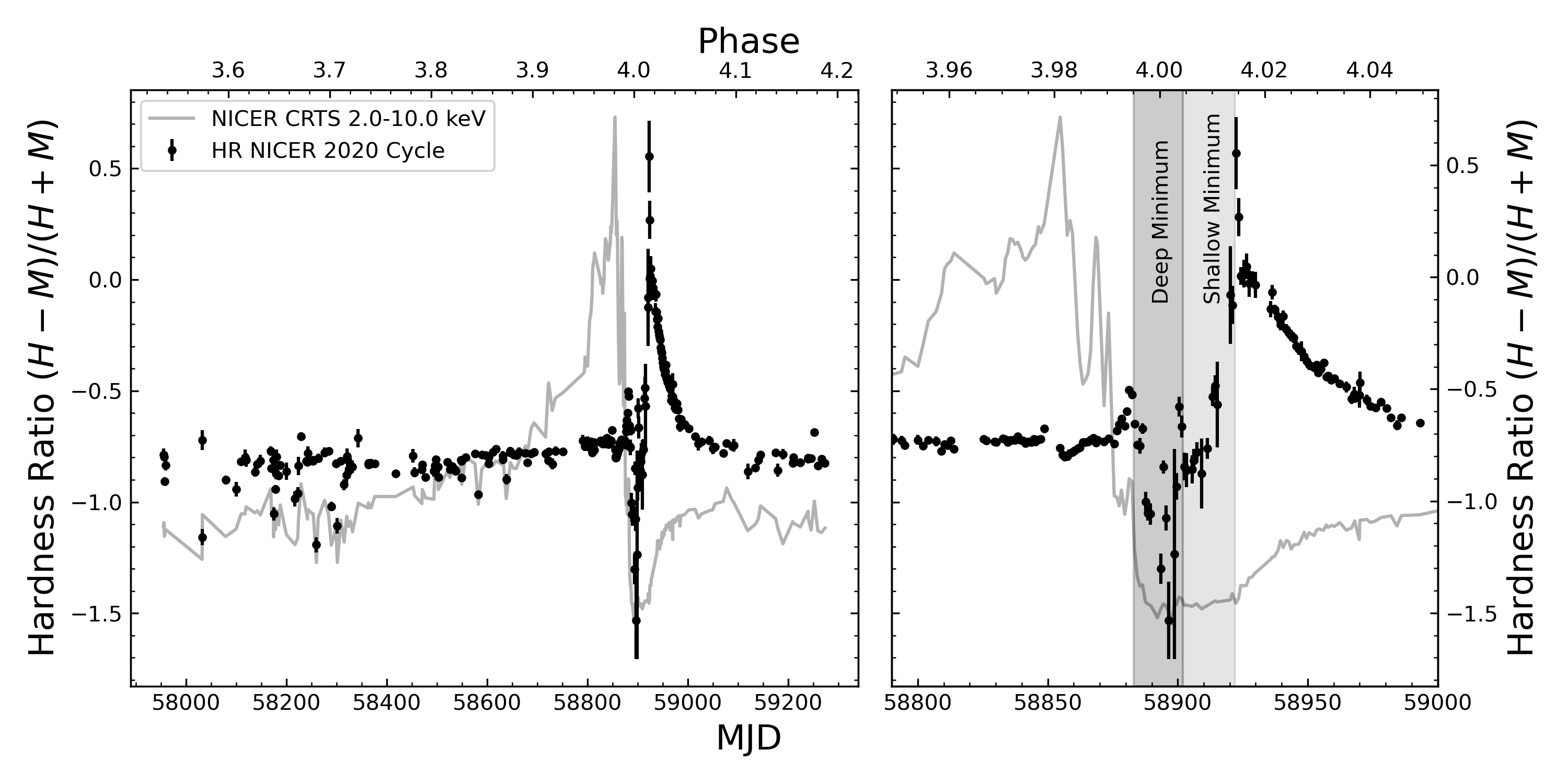}
  \caption{Hardness ratio observed by \nicer\ using a medium band between 2.5 - 3.5 keV and a high band between 6.5 - 7.5 keV.
    The gray regions show the boundaries of the ``deep'' and ``shallow'' X-ray minimum.
    The abrupt decline in hardness ratio marks the start of the deep minimum.
    The abrupt rise in hardness ratio marks the end of the deep minimum and start of the shallow minimum.
    The shallow minimum interval ends when the 2-10 keV flux begins to noticeably increase and the observed hardness is a maximum, indicating the re-emergence of highly absorbed emission from the CWR.
 }
  \label{fig:eta_car_nicer_hr}
\end{figure}

Figure \ref{fig:eta_car_nicer_hr} shows the derived HR from the \nicer\ observations.
The HR shows a nearly linear increase with a small positive slope from mid-cycle near apastron to about 20 days before the start of the X-ray minimum, indicating that the hard band flux increases more rapidly compared to the soft band flux, probably due to residual absorption in the soft band.
After this, there is a significant increase in the HR occurring in a short phase interval, $\Delta\phi\approx 0.005$~days, due to a combined increase in emission measure of the shocked wind from the secondary and increased absorption by the wind of \ec-A as the colliding-wind shock starts to move closer to and behind the primary star.
This hardness increase happens at the same time the colliding-wind flux is at its observed maximum just before the plunge to minimum.
When the plunge to flux minimum starts, the HR also decreases, implying the higher energy emission is decreasing rapidly compared to lower-energy emission.  
This abrupt decline marks the start of the ``deep'' X-ray  minimum, when the source spectrum is dominated by soft emission from the ``Central Constant Emission'' (CCE) component \citep{2007ApJ...663..522H}, cosmic background in the \nicer\ field-of-view, and uncertainties in correction of background contamination, and the colliding-wind source emission is depressed.

Shortly thereafter, the HR abruptly increases marking the end of the deep minimum, followed by a short interval during which the HR continues to increase, but at a slower rate.
This time corresponds to the ``shallow'' part of the X-ray minimum.
As the colliding-wind source strengthens, the HR abruptly reaches a maximum (which corresponds to the end of the post-minimum flux increase in the 6.5~--~7.5~keV band shown in Figure~\ref{fig:eta_car_nicer_crts}), then declines in a quasi-exponential fashion, as the colliding-wind emission increases but the soft band emission is still heavily absorbed by the wind of \ec-A.

\section{The \nicer\ X-ray Spectrum}
\label{sec:spec}

\begin{figure}[htbp]
  \centering
  \includegraphics[width=\linewidth]{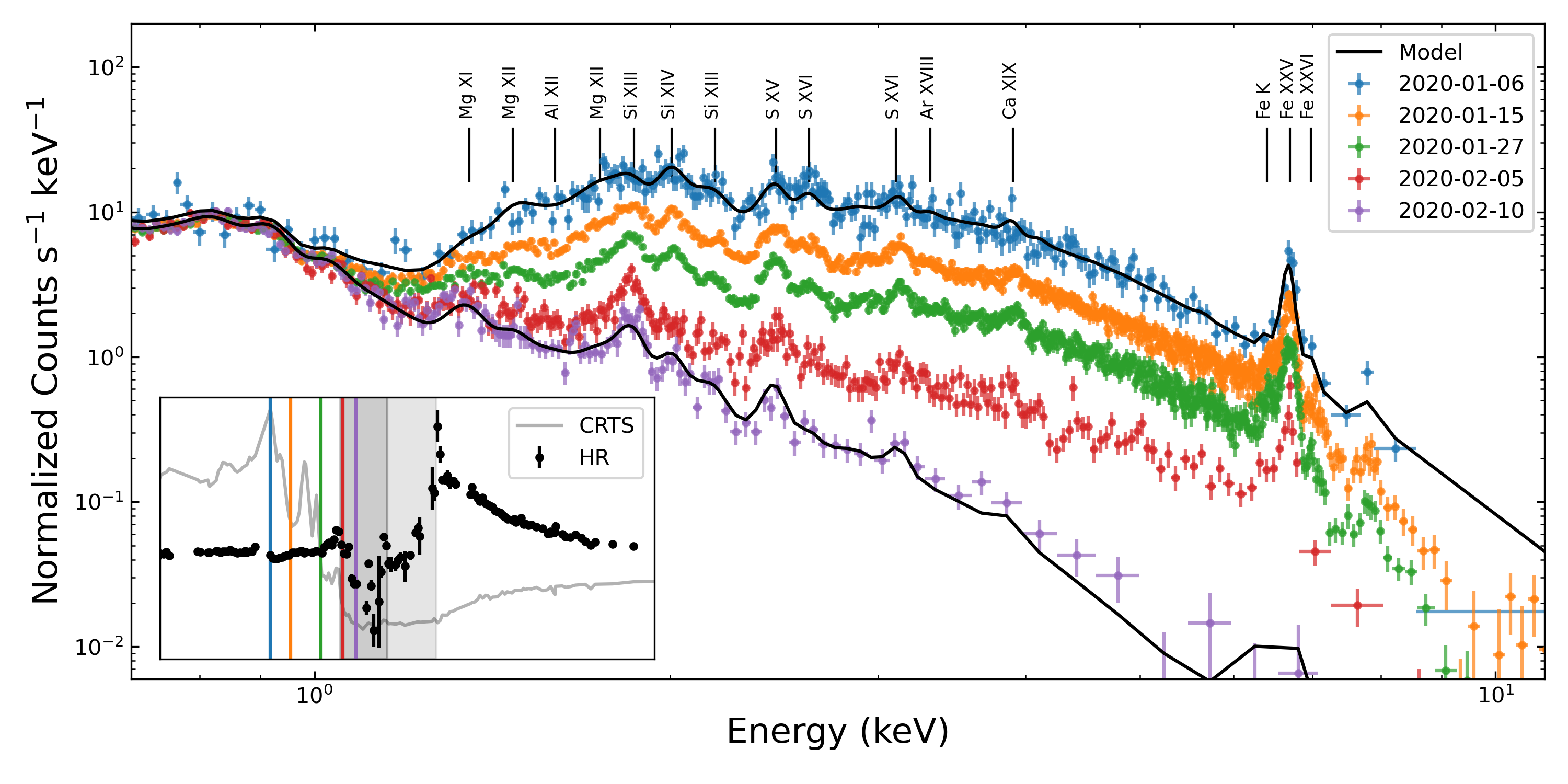}
  \caption{\ec\ spectrum at different phases during the X-ray plunge.
    The plot has different spectra before the last X-ray minimum showing the decrease of the spectrum.
    The inset shows HR in black and count rates in gray against time.
    The color code lines show the time when the spectrum was observed and the gray regions are deep and shallow minimum respectively.
    We can see the changes above 1.0 keV.
    The three upper spectra show a evenly distributed decline in count rates at energies $>$ 1.0 keV.
    By the moment of the fourth observation in red we start to notice a decline in mid range energies between 1.5 - 4.0 keV.
    The last spectrum it is 7 days before the X-ray minimum, noticed the disappearing of the high energy band above 4.0 keV.
  }
  \label{fig:eta_car_nicer_spectrum_down}
\end{figure}

Figure \ref{fig:eta_car_nicer_spectrum_down} shows \nicer\ spectra from the last X-ray maximum (06~January~2020), through the plunge, and until the last X-ray minimum (10~February~2020). 
Figure \ref{fig:eta_car_nicer_spectrum_down} is similar to Figure 2 in \cite{2021ApJ...914...47K} but chosen to avoid flare peaks.
Some strong emission lines are marked.
The first three spectra show a gradual decline in count rate above 1.5 keV.
By 5 February the harder X-ray emission near 1.5 keV is so low that the CCE is clearly observed in the region close to 2.0~keV.
The strong Fe XXV feature is still present on 5 February, indicating that hard X-ray emission disappears completely only at minimum.

\begin{figure}[htbp]
  \centering
  \includegraphics[width=\linewidth]{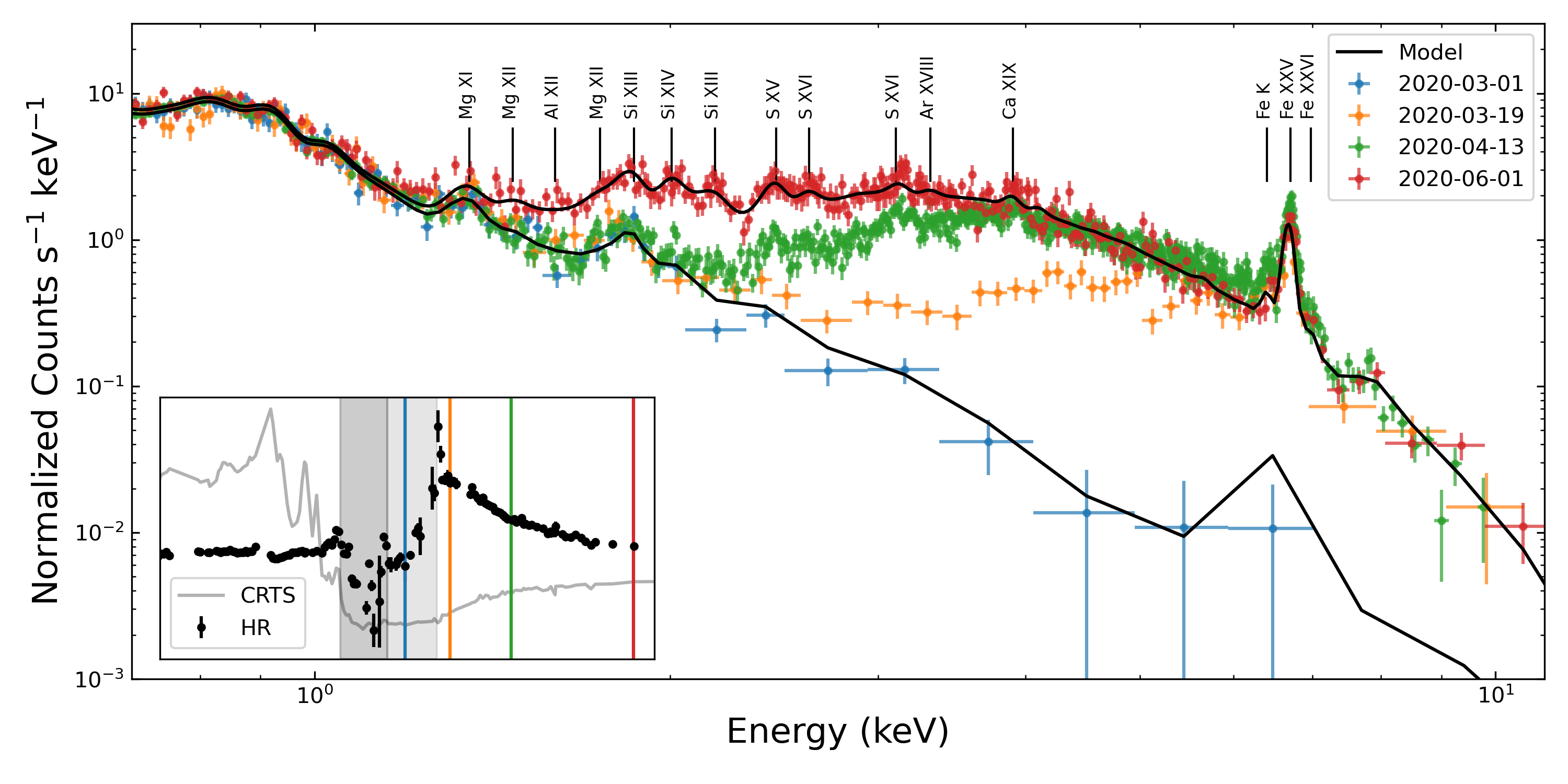}
  \caption{\ec\ spectrum at different phases after the X-ray minimum.
    The plot has different spectra after the last X-ray minimum showing the recovery of the spectrum.
    As in Figure \ref{fig:eta_car_nicer_spectrum_down}, here we can follow the changes above 1.0 keV during the recovery, and inset with the same color code as the spectra shows HR and count rates against time.
    The blue spectrum is in the shallow minimum and does not show emission above 4.0 keV.
    By the time \ec's X-ray emission is out of the shallow minimum the high X-ray emission above 4.0 keV is recovered but the mid range energies are highly affected by column density (See \ref{fig:eta_car_nicer_nh}) which here is reflected in the HR on the inset plot.
    In the last two spectra (green and red), we can see how the mid range X-ray emission continue recovering slowly after months of the X-ray minimum, but the emission $>$ 4.0 keV was completely recovered since the exit of the shallow minimum.
  }
  \label{fig:eta_car_nicer_spectrum}
\end{figure}

Figure \ref{fig:eta_car_nicer_spectrum} shows a sample of  four \nicer\ spectra of \ec\ from the X-ray minimum (01~Mar~2020, in blue) up to three months later after the X-ray emission has completely recovered from the X-ray minimum (1~Jun~2020, in blue).
Figure \ref{fig:eta_car_nicer_spectrum}  can be compared to Figure~4 from \cite{2007ApJ...663..522H}, which shows a similar montage of the \xmm\ spectra of \ec\ from 2003.
The colliding wind emission cannot be seen below 1.5 keV due to absorption from the Homunculus nebula, which correspond to N$_{H}\sim 3\times$10$^{22}$ cm$^{-2}$.
The emission below 1.0 keV comes from the OE and does not change significantly.
The emission above 4.0 keV recovers faster than the emission between 2.0 - 4.0 keV since it is less affected by absorption, as can be seen by comparing the 19 March and 1 March spectra.
By 3~April~2020 ($\sim$30 days after the start of the X-ray minimum)  the flux above 4~keV has fully recovered, with the colliding wind flux below 4~keV continuing to increase through June 1.
Emission above 1.5 keV is mostly thermal emission from the hot shocked wind from the companion star, \ec-B, while below 1.5~keV emission is dominated by the soft thermal emission in the OE.
It is worth noting that the Fe~XXV line can be detected by \nicer\ even in a short (889~s) observation during the X-ray minimum. In addition to the thermal line emission, \nicer\ clearly shows the Fe-K fluorescence line at 6.4~keV \citep[See][]{1998ApJ...494..381C} blended with the 6.7~keV Fe~XXV triplet.   

\subsection{X-ray Template Model}\label{sec:x-ray_template}

\ec's spectrum varies over different phases, changing the physical parameters that produce the X-ray emission from the CWR.
In order to obtain a better estimation of the temperatures and the column densities from the \nicer\ spectrum we used an \ec\ spectrum template using all available dispersed spectra from the \chandra\ HETG \citep[][Espinoza-Galeas et al. 2022, in prep, ]{2020AAS...23537704E,2021PhDT.........5E}.
We combined the 32 \chandra\ HETG observations in Table \ref{tab:chandra_obs} to obtain a high-precision dispersed X-ray spectrum with a total exposure of 2~Msec, with a exposure weighted average phase of 0.73.  
This is the highest precision, highest spectral resolution spectrum of \ec\ currently available.
We fit the combined spectrum  with a linear combination of  a small number of collisionally-ionized plasma models\footnote{http://atomdb. org/} plus a Gaussian line to model the  fluorescence Fe~K emission line, plus absorption \citep{2000ApJ...542..914W}.
The \nicer\ spectra are not sensitive to the broadening of the emission lines, but using the \chandra\ combined spectrum it was clear that a velocity-broadened model was necessary.
We found that two velocity-broadened non-solar abundance thermal components were sufficient to describe the combined HETG spectrum.
We used the \chandra\ combined spectrum as a template model to fit each \nicer\ spectrum to derive X-ray spectral parameters.
We fixed the abundances and line broadening to the \chandra\ values when fitting the \nicer\ spectra.

\begin{deluxetable*}{rrrrr}
\tabletypesize{\scriptsize}
%\tablecolumns{7}
%\tablewidth{0pc}
\tablecaption{HETG \chandra\ observations \label{tab:chandra_obs}}
\tablehead{\colhead{obsid} & \colhead{expt} & \colhead{Date} & \colhead{MJD} & \colhead{Phase}}
%\colhead{} &	\colhead{} &	\colhead{} &	\colhead{} &	\colhead{} &	\colhead{} &	\colhead{}\\
%\colhead{} &	\colhead{} &	\colhead{} &	\colhead{} &	\colhead{} &	\colhead{} &	\colhead{}\\
%}
\startdata
632   &	89545.68 &	2000-11-19T02:47:43 &	51867.65 &	0.527 \\
3749  &	91280.92 &	2002-10-16T08:09:53 &	52563.89 &	0.871 \\
3745  &	94533.00 &	2003-05-02T11:57:20 &	52762.07 &	0.969 \\
3748  &	97249.20 &	2003-06-16T05:36:31 &	52806.82 &	0.991 \\
%3746  &	90310.91 &	2003-07-20T01:47:26 &	52840.62 &	1.008 \\
%3747  &	70106.47 &	2003-09-26T22:46:56 &	52909.38 &	1.042 \\
7445  &	25393.24 &	2008-10-15T20:18:41 &	54755.01 &	1.954 \\
10787 &	68904.23 &	2008-10-21T23:11:26 &	54761.38 &	1.957 \\
10831 &	17589.25 &	2008-12-08T12:31:41 &	54808.63 &	1.980 \\
8930  &	29646.37 &	2008-12-10T01:49:02 &	54810.26 &	1.981 \\
10827 &	27365.77 &	2008-12-12T17:34:07 &	54812.90 &	1.983 \\
10895 &	15257.55 &	2009-03-16T17:18:12 &	54906.82 &	2.029 \\
10894 &	21986.45 &	2009-03-17T21:26:40 &	54908.04 &	2.030 \\
9945  &	31275.61 &	2009-04-21T06:46:35 &	54942.48 &	2.047 \\
10905 &	26298.47 &	2009-04-26T13:56:49 &	54947.75 &	2.049 \\
9946 & 56628.15  &      2009-09-06T20:45:21 &   55081.21 &      2.116 \\
11992 &	18428.78 &	2009-12-21T12:52:08 &	55186.66 &	2.167 \\
11017 &	17454.93 &	2009-12-22T09:47:25 &	55187.52 &	2.168 \\
12064 &	17703.14 &	2009-12-23T03:48:06 &	55188.28 &	2.168 \\
12065 &	18625.89 &	2009-12-23T23:44:42 &	55189.11 &	2.168 \\
11993 &	43826.40 &	2010-11-14T13:33:42 &	55514.84 &	2.329 \\
11994 &	39364.24 &	2010-11-21T07:26:40 &	55521.56 &	2.333 \\
12358 &	102189.1 &	2011-10-20T11:19:40 &	55855.10 &	2.497 \\
13670 &	31173.59 &	2012-10-19T16:25:08 &	56219.88 &	2.678 \\
15569 &	68162.57 &	2012-10-20T17:29:41 &	56221.15 &	2.678 \\
21177 &	29027.65 &	2019-05-16T02:58:19 &	58619.31 &	3.863 \\
22218 &	15446.14 &	2019-05-17T00:32:15 &	58620.12 &	3.864 \\
22219 &	33862.69 &	2019-05-18T17:35:26 &	58621.94 &	3.864 \\
21178 &	29024.17 &	2019-07-01T17:56:35 &	58665.93 &	3.886 \\
22272 &	27089.92 &	2019-07-02T10:13:04 &	58666.60 &	3.887 \\
22273 &	15489.07 &	2019-07-04T10:01:32 &	58668.52 &	3.887 \\
21179 &	57150.21 &	2019-08-27T05:54:31 &	58721.71 &	3.914 \\
21180 & 24249.47 &      2019-10-08T13:56:11 &   58764.74 &      3.937 \\
22310 & 19570.27 &      2020-01-06T01:58:12 &   58854.21 &      3.981 \\
23117 & 19570.34 &      2020-01-06T17:13:59 &   58854.84 &      3.981 \\
23119 & 19567.59 &      2020-01-11T06:06:16 &   58859.38 &      3.983 \\
22311 & 34169.62 &      2020-01-13T11:07:07 &   58861.68 &      3.984 \\
23126 & 34168.16 &      2020-01-14T10:03:59 &   58862.63 &      3.985 \\
22846 & 29302.70 &      2020-01-16T02:34:15 &   58864.29 &      3.986 \\
22847 & 14703.70 &      2020-01-26T16:22:24 &   58874.78 &      3.991 \\
23131 & 33195.76 &      2020-01-27T16:42:56 &   58875.90 &      3.991 	
\enddata
\end{deluxetable*}

To model the \nicer\ spectra, we also added to the HETG model a third non-variable thermal component to account for non-variable soft emission below 1.5 keV from the shocked gas in the OE (which is not strongly visible in the HETG spectrum).
The initial spectral model parameters are given in Table~\ref{tab:nicer_model}.
We then fit all the net \nicer\ spectra, allowing temperatures, column densities and emission measures (normalization) of the two colliding-wind components to vary, with other components and the OE emission component held fixed.  

A complete detailed analysis of the combined HETG \chandra\ spectrum is in \cite{2021PhDT.........5E}.
To compare \nicer\ temperatures and column densities we also present preliminary results of temperatures and column densities using the individual HETG \chandra\ spectra (Espinoza-Galeas et al. 2022, in prep).

\begin{deluxetable}{rrrrr}\label{tab:nicer_model}
\tabletypesize{\scriptsize}
\tablecolumns{4}
\tablewidth{0pc}
%\tablecaption{\nicer\ initial model used to fit \ec's spectrum. The model is {\tt TBabs*vapec + TBabs*vapec + gaussian}}
\tablecaption{Initial Spectrum Model}
\tablehead{
  %\colhead{} &	\colhead{Colliding Wind Components} & \colhead{} & \colhead{}\\
  \colhead{Parameter} & \colhead{Cooler Comp} & \colhead{Hotter Comp} & \colhead{Outer Ejecta Comp}  
  }
\startdata
% \hline
% \multicolumn{3}{c}{\textit{Colliding Wind Components}} \\
\hline
 NH  ($10^{22}$~cm$^{-2}$ &    3.00      &  10.00    &    0.60                    \\
 kT    (keV)              &    1.26      &  4.43     &    0.25                     \\
 Redshift                 &    0.00      &  0.00     &    0.00                   \\
 Broadening (km s$^{-1})$    &     598.00      &  888.88    &    900.00                       \\
 %norm       &     1.00         &  1.00    &    1.00    & ---             \\
 He        &     1.00    & fixed        & fixed                  \\
 C          &     1.00    & fixed        & fixed                  \\
 N          &    20.00     & fixed        & fixed                  \\
 O          &     1.00     & fixed        & fixed                   \\
 Ne         &     1.00     & fixed        & fixed                  \\
% Redshift   &     1.88533E-03                    \\
%  \multicolumn{3}{c}{\textit{Common Fixed Abundances}} \\
 Mg         &     0.59    & fixed        & fixed            \\
 Al         &     0.47    & fixed        & fixed            \\
 Si         &     0.41    & fixed        & fixed           \\
 S          &     0.46    & fixed        & fixed           \\
 Ar         &     0.66    & fixed        & fixed           \\
 Ca         &     0.88    & fixed        & fixed           \\
 Fe         &     0.51    & fixed        & fixed          \\
 Ni         &     1.00     & fixed        & fixed          \\
%% \multicolumn{3}{c}{\textit{Fe K Fluorescence Line}} \\
% LineE      & ---             & ---          & ---       &    6.40                            \\
% Sigma      & ---             & ---          & ---       &    1.00                        \\
 \hline
% \multicolumn{3}{c}{\textit{Outer Ejecta}} \\
%\hline
% nH         &    0.600000                            \\
% kT         &     0.250000                           \\
% Mg         &     0.590000                           \\
% Al         &     0.470000                           \\
% Si         &     0.410000                           \\
% S          &     0.460000                           \\
% Ar         &     0.660000                           \\
% Ca         &     0.880000                           \\
% Fe         &     0.510000                           \\
% Ni         &     1.00000                            \\
% Redshift   &     1.13800E-02                        \\
% Velocity   &     0.0                                \\
 %norm       &     0.0                                \\
\enddata
\tablecomments{Elemental abundances are relative to solar values, using \cite{1989GeCoA..53..197A}. The model also includes a variable gaussian iron K line near 6.4 keV.}
\end{deluxetable}            

%@ARTICLE{1989GeCoA..53..197A,
%       author = {{Anders}, E. and {Grevesse}, N.},
%        title = "{Abundances of the elements: Meteoritic and solar}",
%      journal = {\gca},
%     keywords = {Abundance, Chemical Analysis, Chondrites, Meteoritic Composition, Photosphere, Solar Corona, Energetic Particles, Isotopes, Solar Wind, METEORITES, ABUNDANCE, SAMPLES, METEORITE, CI CHONDRITES, SUN, CORONA, PHOTOSPHERE, REVIEW, ELEMENTS, FRACTIONATION, SOLAR SYSTEM, NUCLIDES, SOLAR WIND, CHARGED PARTICLES, RARE GASES, REFRACTORY ELEMENTS, COMETS, SIDEROPHILES, VOLATILES, NUCLEOSYNTHESIS, HALLEY, COMPARISONS, Astrophysics; Meteorites},
%         year = 1989,
%        month = jan,
%       volume = {53},
%       number = {1},
%        pages = {197-214},
%          doi = {10.1016/0016-7037(89)90286-X},
%       adsurl = {https://ui.adsabs.harvard.edu/abs/1989GeCoA..53..197A},
%      adsnote = {Provided by the SAO/NASA Astrophysics Data System}
%}

\subsection{Flux Variations}

\begin{figure}[htbp]
  \centering
  \includegraphics[width=\linewidth]{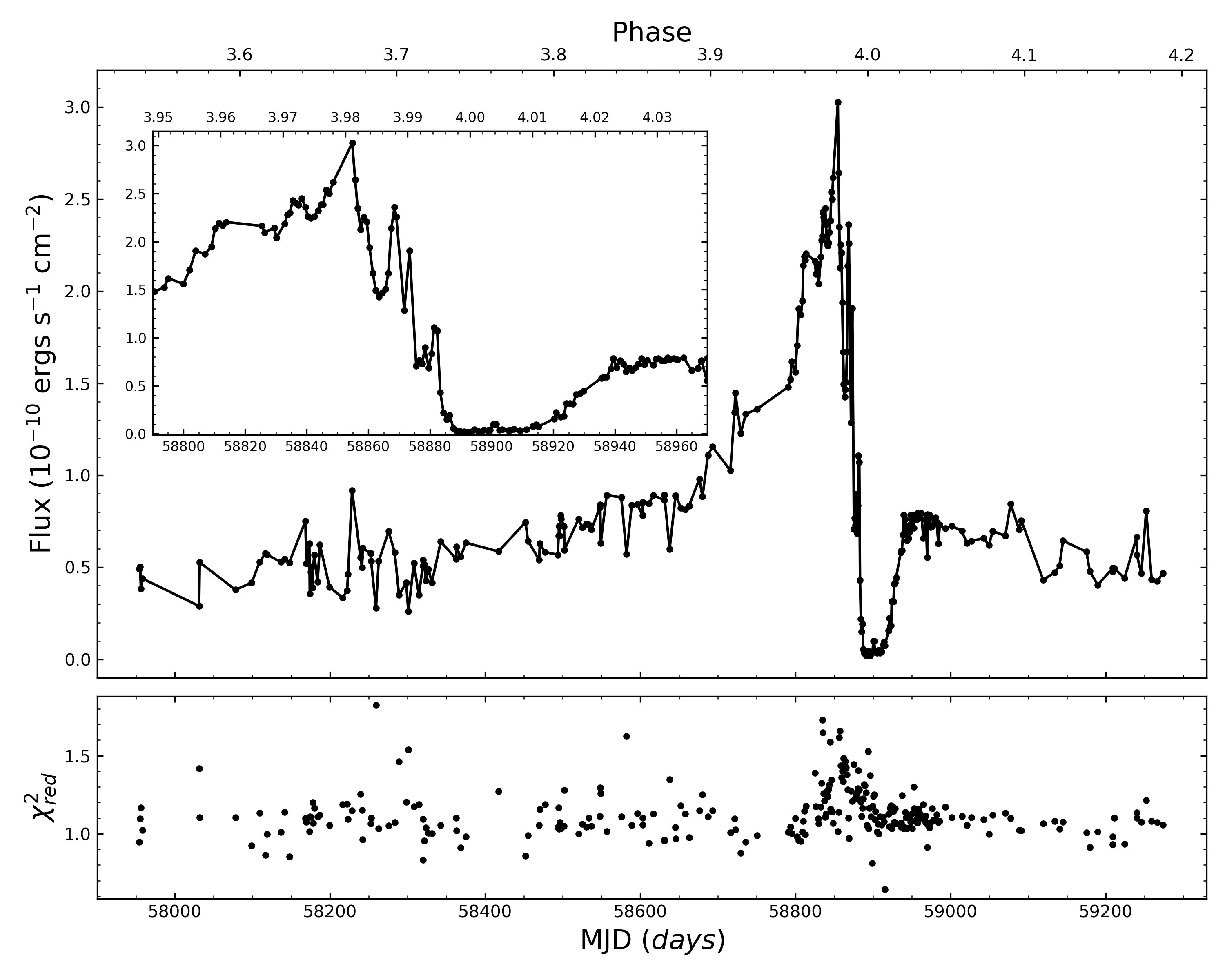}
  \caption{Flux between 2.0 - 10.0 keV observed by \nicer.
    Fluxes are measured in {\tt XSPEC} command {\tt Flux} fitting a three collisional ionized plasma model with a Gaussian for the Fe K line.
    The bottom panel shows the reduced $\chi^{2}$ from the fittings.
    At this point in our analysis, the background subtraction becomes more important since the goodness of the fit is highly affected if the background is not properly subtracted. 
    Both background estimators improve the goodness of the fit but the 3C50 estimator shows a more stable behavior over the whole data set of observations.
  }
  \label{fig:eta_car_nicer_flux}
\end{figure}

Figure \ref{fig:eta_car_nicer_flux} shows the \nicer\ flux between 2.0~--~10.0 keV of all the \nicer\ observations using the \texttt{nibackgen3C50} estimator to correct for background, although this model overestimates background for a small number of observations under conditions of high background. 
The bottom panel shows the reduced $\chi^{2}$ value for all the fits, which is generally acceptable except close to X-ray minimum when uncertainties in background subtraction play a significant role in defining the net spectrum.
The inset highlights the variations near the X-ray minimum.  

\begin{figure}[htbp] %  placement: here, top, bottom, or page
   \centering
   \includegraphics[width=\linewidth]{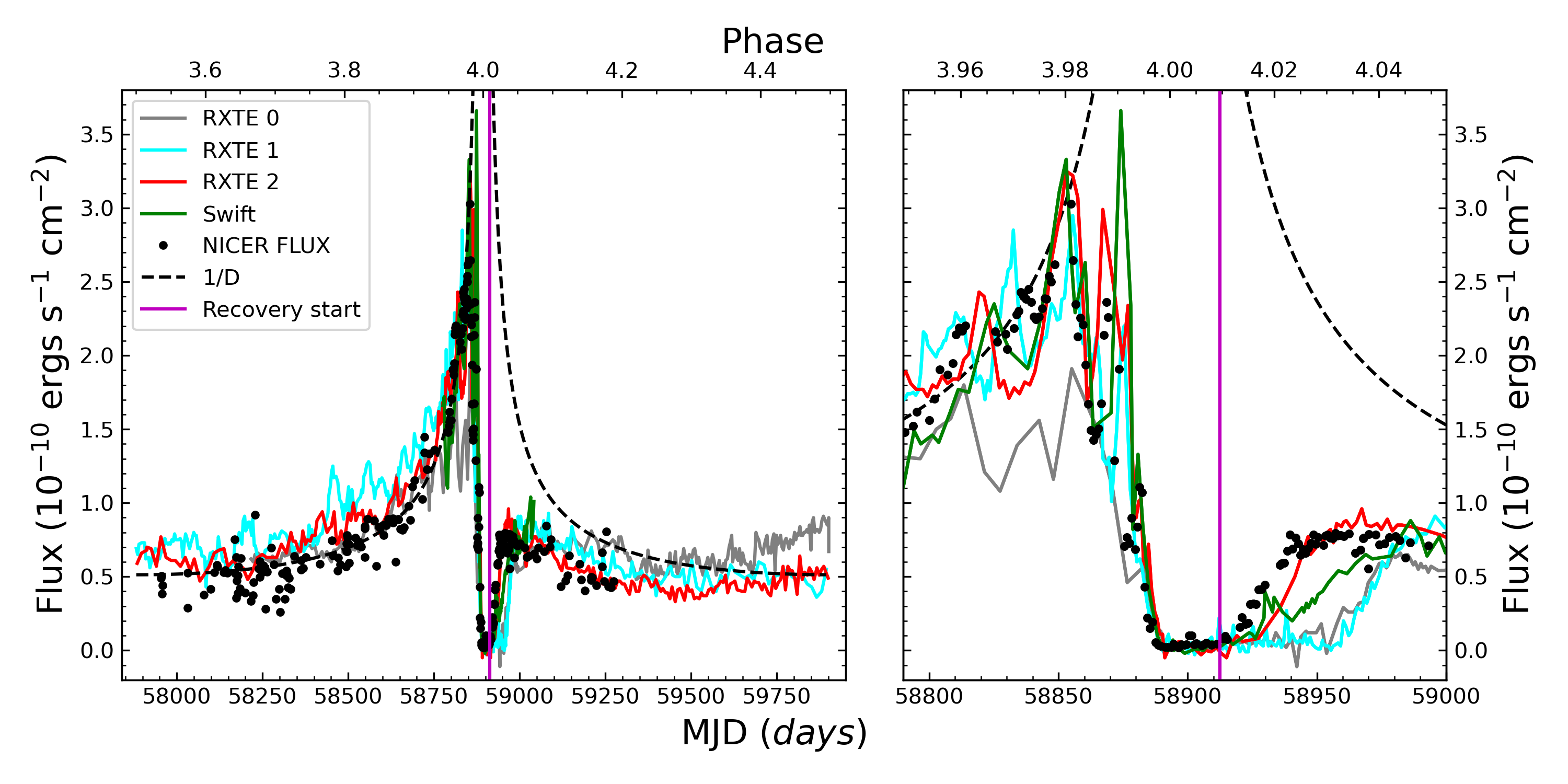} 
   \includegraphics[width=\linewidth]{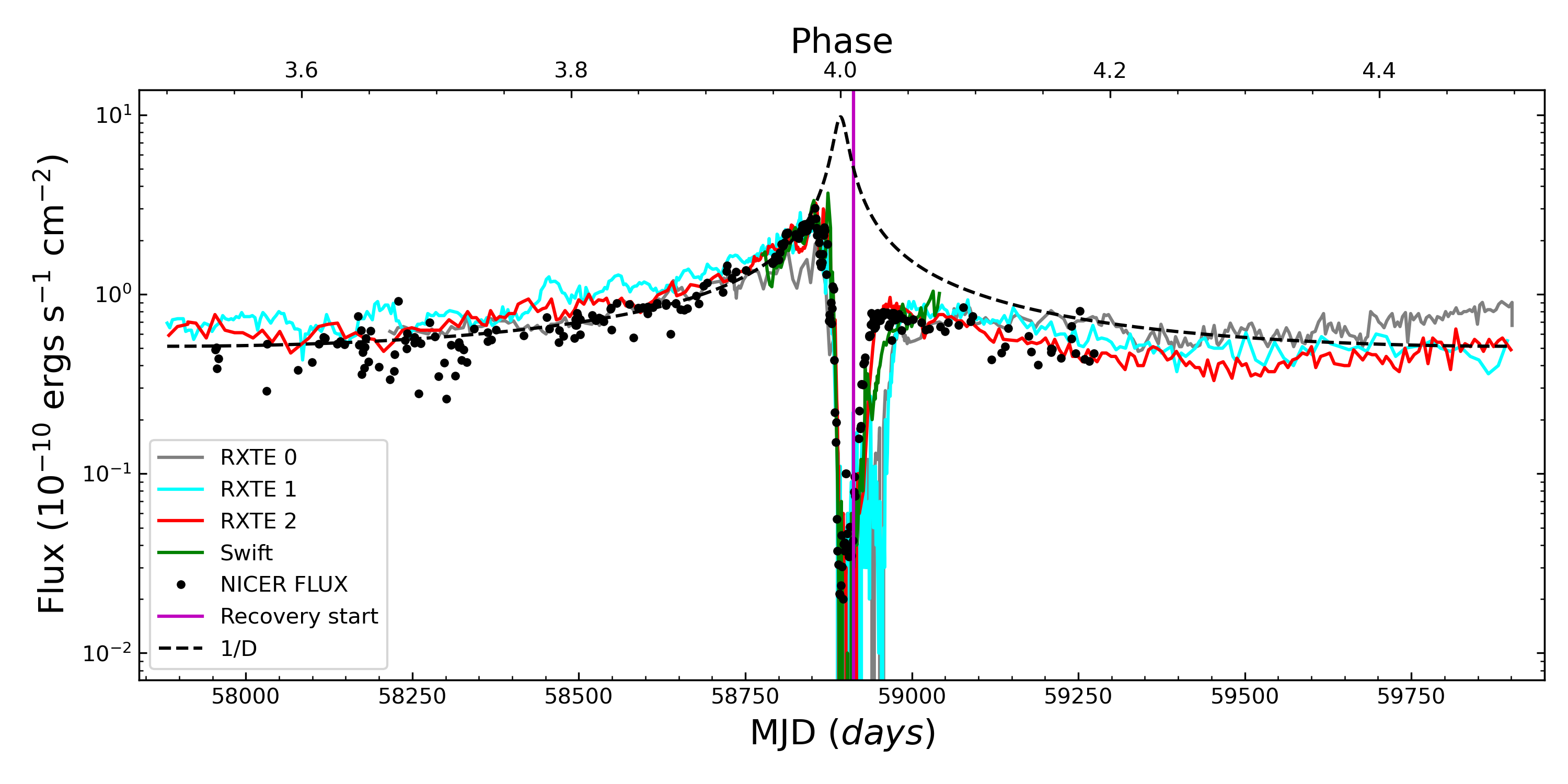} 
   \caption{\textit{Top}: Comparison of the  2~--~10~keV band fluxes from \nicer, and earlier \rxte\ and \swift\ observations from \cite{2017ApJ...838...45C}. The earlier measures have been advanced by 4, 3, 2, and 1 periods for the \rxte\ cycle 0, 1, 2 and \swift\ data, respectively, for comparison to the \nicer\ measures. A $1/D$ curve is shown by the dashed line. \textit{Bottom}: Flux in $\log$ scale, to emphasize the depth of the X-ray minimum and the maximum height of the $1/D$ curve.}
   \label{fig:rxteswiftnicer}
\end{figure}

Figure \ref{fig:rxteswiftnicer} shows the flux between 2.0 -- 10.0 keV in $10^{-10}$ \fluxcgs\ compared with \rxte\ and \swift\ fluxes in the same band \citep[taken from][]{2017ApJ...838...45C}.
The three telescopes have instrumental differences and different fields of view that can cause differences in the flux in the same range due to the varying amount of cosmic and instrumental background contamination and other factors.
We calibrated the data from \nicer\ with the \rxte\ and \swift\ data by subtracting a small amount of flux from each flux measure to match the flux of the X-ray deep minimum in  all the cycles.
During the deep minimum the net flux is close to the cosmic X-ray background and  flux differences are minimal then.

X-ray flux from a colliding-wind shock in an eccentric binary should vary inversely with the separation $D$ of the stars \citep{usov92, Stevens:1992yu} if the shock cools adiabatically.
The smooth dashed curve in Figure \ref{fig:rxteswiftnicer} shows a $1/D$ variation using the orbital elements in Table~\ref{tab:sysparams}.
For most of the orbit, the \nicer\ 2~--~10 keV X-ray flux agrees with the $1/D$ curve, but starts to deviate from it 10 days before the X-ray minimum is reached.
Prior to X-ray minimum, the \rxte\ fluxes lie on the  $1/D$ curve which fits the \nicer\ data, though  the agreement is better about 80 days after X-ray minimum.

Once reaching maximum flux, the \nicer\ lightcurve shows the same plunge into the X-ray minimum as seen in the earlier orbital cycles, as shown in the inset in the upper panel in Figure~\ref{fig:rxteswiftnicer}.
The X-ray minimum is caused by a combination of eclipse and the disruption of  the shock around \ec~B by the wind of \ec~A, as discussed by \cite{Hamaguchi:2014lr}.
According to Eq.~\ref{eq:ephem}, the deep X-ray minimum should have been reached around 14 February 2020. \nicer\ observed the deep minimum around 13 February 2020 as we reported in \cite{2020ATel13516....1C}.

Based on the behavior shown in previous cycles, the end of the X-ray minimum was expected no earlier than the end of March.
But  the \nicer\ data show a clear increase in flux between 0.5 - 9.0 keV starting approximately on 15 March 2020 \citep[MJD 58923.0, ][]{2020ATel13636....1E}. 
This was the earliest recovery of \ec's X-ray emission yet observed. The flux increase from minimum in the 2~--~10~keV band ended by MJD 58950 (2020-04-11). The analogous analysis in \cite{2021ApJ...914...47K} shows similar results, confirming our announcement in \cite{2020ATel13636....1E}.

\subsection{The X-ray Period}

\begin{figure}[htbp] %  figure placement: here, top, bottom, or page
   \centering
   \includegraphics[width=1.0\linewidth]{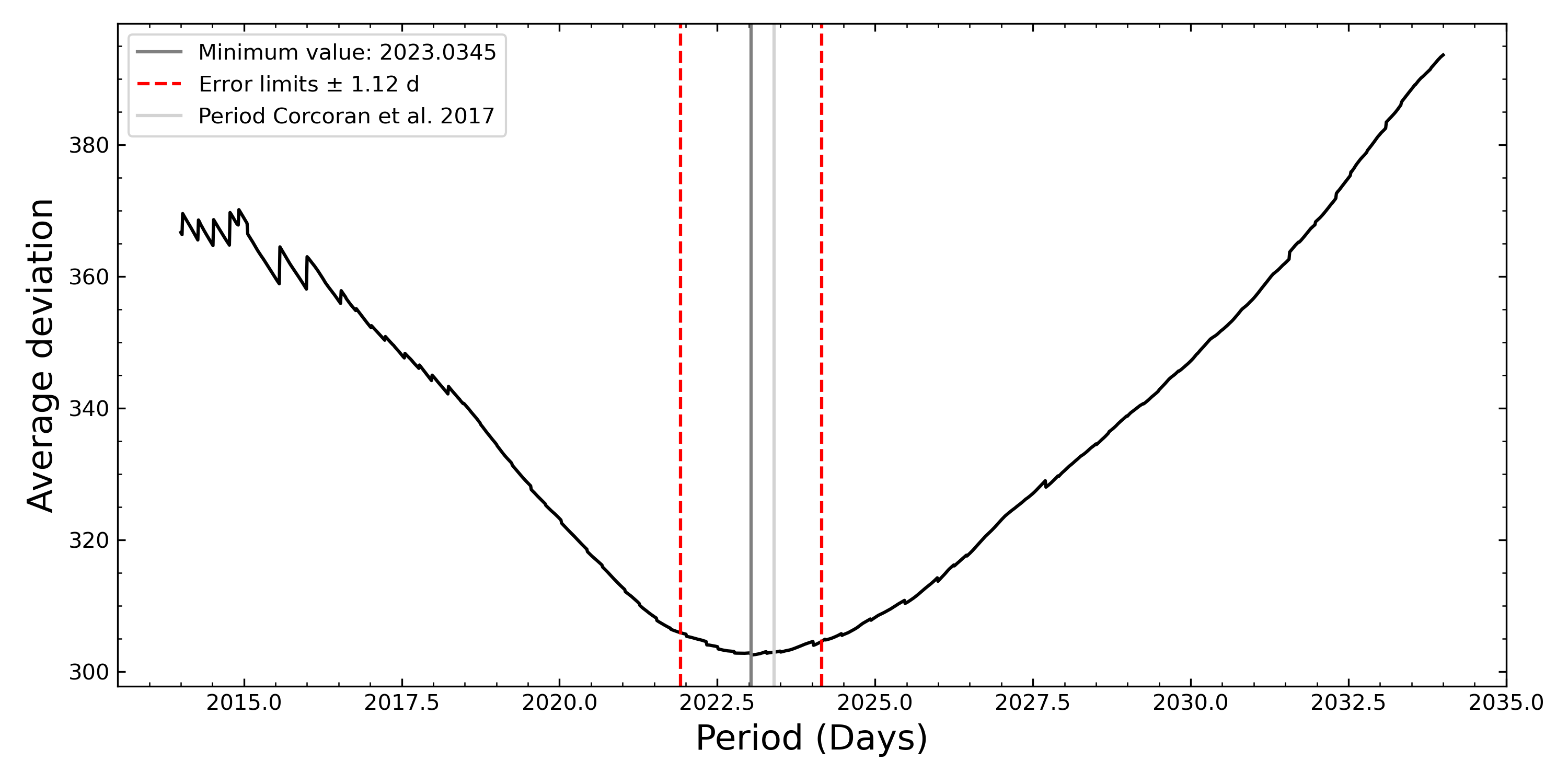} 
   \caption{Sum of the square of the residuals for the \rxte, \swift, and \nicer\ X-ray flux curves phased for periods in the range 2015~--~2035 days. 
     Including the \nicer\ data, the best X-ray period is 2023.035 days (dark gray line).
     The light gray line is the period calculated in \cite{2017ApJ...838...45C}.
     The red lines are the limits of the error calculated in this work.
   }
   \label{fig:period}
\end{figure}

We re-determined the X-ray period including the \nicer\ data using a simple implementation of a phase-dispersion minimization technique \citep{1978ApJ...224..953S}.
We phase-folded the four X-ray flux curves from  \rxte, \swift \citep{2017ApJ...838...45C}, and \nicer\ using trial periods in the range 2015~--~2025 days.
For each trial period, we interpolated each flux curve to a common phase scale assuming a common epoch, then calculated the sum of the residuals between the three flux curves relative to the \nicer\ fluxes.
Figure~\ref{fig:period} plots the summed residuals versus test phase.
We found that a period of 2023.03 $\pm$ 1.12 days yielded the smallest summed residuals.
This period is 0.36 days shorter than the X-ray period derived by \cite{2017ApJ...838...45C}, and 0.33 days longer than the period derived from analysis of the He~II~$\lambda$4686 emission line by \cite{2016ApJ...819..131T}.

\cite{2017ApJ...838...45C} calculated an error of 0.71 which is smaller than our calculation.
This is probably due to the high variability of the first observations made with high background conditions.
But, the calculation of the period in this work is still in very good agreement with \cite{2016ApJ...819..131T} and \cite{2017ApJ...838...45C}.
We decided to keep the period of 2023.40 $\pm$ 0.71 days to try to avoid the uncertainty added by the first \nicer\ observations.

\subsection{X-ray Flaring}
\label{sec:Flaring}

Just prior to minimum, the X-ray emission from \ec\ brightens and undergoes a period of rapid variability (``flares'') which have been observed by \rxte, \swift, and now \nicer.
X-ray ``flaring'' seen by \nicer\ was first observed on MJD 58802 \citep[15 November 2019,][]{2019ATel13327....1C}, about 90 days prior to the X-ray minimum, similar to the start of the flare interval seen by \rxte\ and \swift. 
During the flare interval, the X-ray flux in the 2~--~10 keV band changes dramatically on timescales of days.
This rapid X-ray variability has been seen by all three instruments.
However, the \rxte\ cycle 0 observations were not obtained as frequently as observations in latter cycles, which meant some of the short-timescale variations were not sampled as completely as they were in subsequent cycles.
Except for the undersampled \rxte\ cycle 0 observations, the maximum fluxes in the flares are similar.

\begin{figure}[htbp] %  figure placement: here, top, bottom, or page
   \centering
   \includegraphics[width=\linewidth]{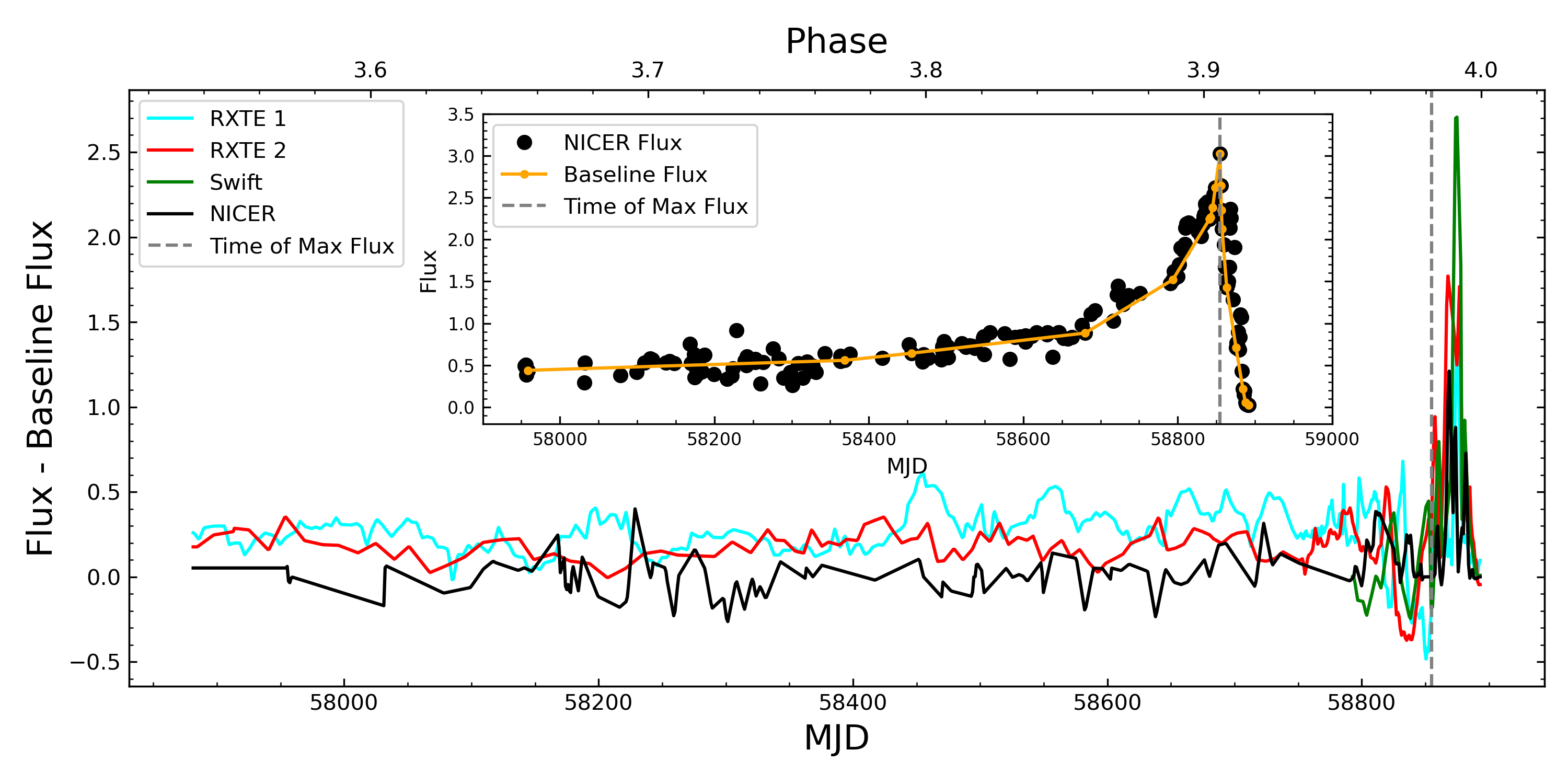}\\ 
   \includegraphics[width=\linewidth]{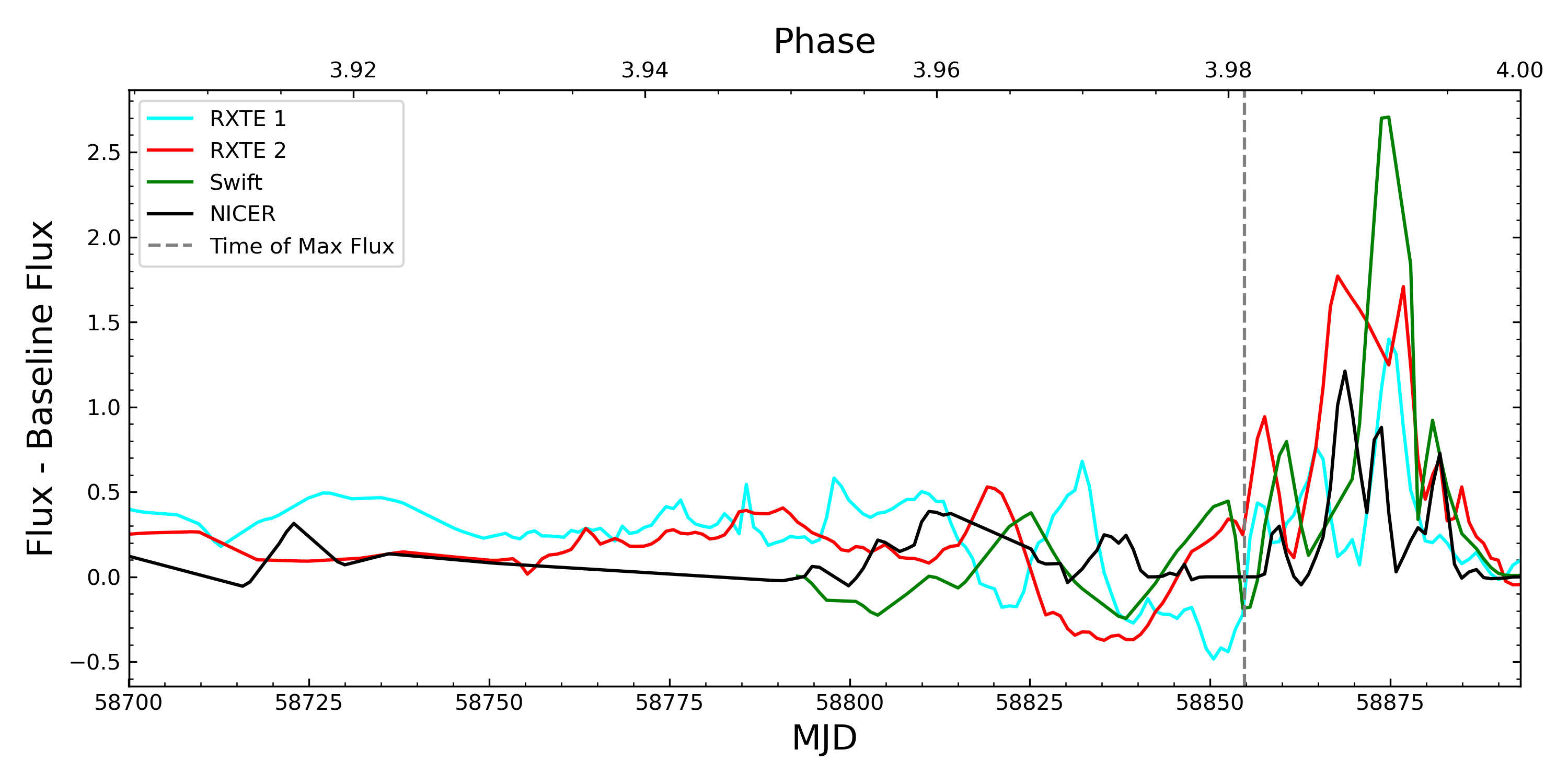} 
   \caption{\textit{Top}: Residuals of \nicer\ (black solid line), \rxte\ (cyan and red), and \swift\ (green). The dashed vertical line shows the observed maximum of the X-ray count rate as seen by \nicer.
   The \nicer\ X-ray 2~--~10 keV flux curve vs. MJD.
   The orange line shows the baseline flux used to calculate the residuals, which approximately follows a $1/D$ variation before maximum, then it is followed by the lowest values during plunge.
   \textit{Bottom}: Same as top zooming the times close to maximum X-ray emission and before minimum X-ray emission. 
 }
   \label{fig:flares_DEG}
\end{figure}

To quantify the flaring observed by \nicer\ we adopted a process similar to that used by \cite{2009ApJ...707..693M}, namely we estimate an underlying smooth flux baseline by eye, and then subtracted this baseline from the net \nicer\ fluxes.
One difference with \cite{2009ApJ...707..693M} analysis is that we use for the baseline flux observations that follow closely the $1/D$ line in Figure \ref{fig:rxteswiftnicer}.
Also we use the same \nicer\ baseline flux to calculate the \rxte\ and \swift\ residuals.
The \nicer\ fluxes, the baseline, and the residuals are shown in Figure \ref{fig:flares_DEG}.
In Figure ~\ref{fig:flares_DEG}, the \nicer\ measures are shown versus time using the epoch of periastron $T_{o}$ given by \cite{2017ApJ...838...45C}.
Figure~\ref{fig:flares_DEG} also compares the residuals for the \rxte\ and \swift\ data (we do not include the first set of \rxte\ observations since the sampling frequency was not as high as in the later two cycles). 

We do note that \cite{2009ApJ...707..693M} argued that long-duration, low-intensity X-ray flares could be seen near the apastron in \rxte\ observations.
But \nicer\ observations near apastron were obtained at the start of the mission and were affected by variations in local charged particle background and scattered solar radiation, so we can not claim that variations before MJD=58308 d ($\sim\phi=$3.71) are signs of intrinsic source X-ray variability.
After $\phi = 3.71$ \ec's \nicer\ observations started to be taken at low background conditions.

After MJD=58400 ($\sim\phi=3.75$), Figure~\ref{fig:flares_DEG} shows small excess seen by \nicer\ during the rise to X-ray maximum.
Similarly, the excess in the residuals for the \rxte\ and \swift\ observations increases after $\phi=3.75$.
At phases $\sim\phi > 3.93$, \rxte\ and \swift\ residuals show a decrease in the width of flares. 
The shortening between flares can be seen by \nicer\ just until $\phi=3.95$ since we do not have observations between $3.92>\phi>3.95$.
The plot shows a break up at $\phi\approx3.97$ when the X-ray flux is approaching its maximum.
The flares are not clearly seen at the same time in each orbital cycle, though some coincidences between cycles do occur.
After X-ray maximum flux, during the plunge, the flares peaks show more coincidences between different cycles but this can be just a bias due to the fast orbital motion close to the periastron passage.

\subsection{Variations in X-ray Absorption}

\begin{figure}[htbp]
  \centering
  \includegraphics[width=\linewidth]{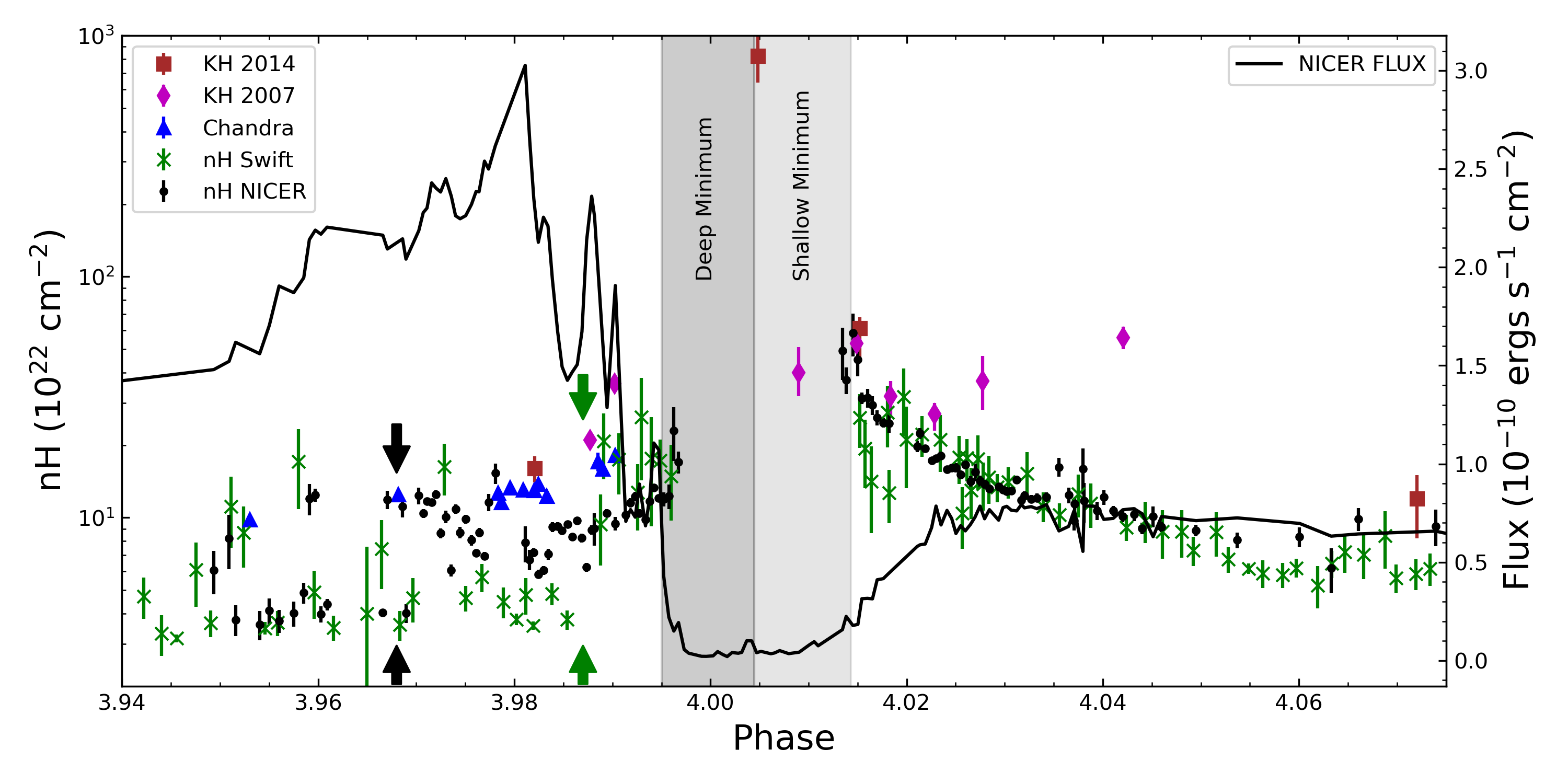}
  \caption{Column densities observed by \nicer\ compared to \swift\ column densities from the last orbital cycle \citep{2017ApJ...838...45C}. 
    The magenta diamonds are previous measurements from \citet[][KH 2007b]{2007ApJ...663..522H}, while the red squares are from \chandra, \xmm, and \suzaku\ spectra \citet[][KH 2014]{Hamaguchi:2014lr}.
    Blue triangles are column densities derived from analysis of \chandra\ grating spectra (D. Espinoza-Galeas et al., 2022, in preparation).  
 The gray regions mark the ``deep'' and ``shallow'' minimum. 
 Notice the good agreement in \swift\ and \nicer\ observations after the shallow minimum.
 The arrows mark the abrupt, step-like increas in column density(near phi=3.99 for Swift and phi = 3.97 for NICER).
}
  \label{fig:eta_car_nicer_nh}
\end{figure}

Figure \ref{fig:eta_car_nicer_nh} shows the derived variation in column density from the analysis of the \nicer\ spectra compared to the column densities derived from the analysis of the \swift\ spectra in 2014 \citep{2017ApJ...838...45C}.
The \nicer\ column densities show substantial scatter between $2\times10^{22}$~cm$^{-2}$ to $2\times10^{23}$~cm$^{-2}$ before phase 3.97.
Derived column densities may be influenced by uncertainties in our estimate of the background and contamination by the soft OE emission below 2 keV.
As flux increases, column-densities measures become more precise and show a quasi-exponential increase reaching a maximum value of $N_{H}\approx 10^{24}$~cm$^{-2}$ the end of the deep X-ray minimum.
Note that near the X-ray minimum determining the column density precisely is difficult because the flux level is so low and the spectrum not well defined in the individual \nicer\ observations.
After reaching maximum, the measured column densities show a quasi-exponential decline through the shallow minimum, which appears to be fairly symmetric to the quasi-exponential increase seen up through the deep minimum.
This is somewhat surprising since the distortion of \ec-A's wind is very different before the X-ray minimum and afterward, as shown by hydrodynamical modeling \citep[see, for example, ][]{Madura:2013fj}.

Figure \ref{fig:eta_car_nicer_nh} also shows previous column density measures from the literature and column densities determined from a new analysis of available \chandra\ High Energy Transmission Grating spectra (Espinoza-Galeas et al., 2022, in prep.).
In general there is good agreement between the \nicer\ column densities and the \swift\ column densities obtained during the previous periastron passage in 2014.
There are significant discrepancies between the \nicer\ column densities and others measured from ``snapshot'' spectra obtained by \suzaku, \xmm, and \chandra\ at some phases.
In particular the maximum column measured by \cite{Hamaguchi:2014lr} is near $10^{25}$cm$^{-2}$, about an order of magnitude larger than the maximum \nicer\ column.
It may be that the maximum \nicer\ column is underestimated due to residual soft circumstellar emission in the \nicer\ field of view and due to uncertainties in background estimation.
There are also large differences, up to an order of magnitude, during the high-absorption interval following X-ray minimum between the \cite{2007ApJ...663..522H} measures, the \chandra\ measure and the \swift\ and \nicer\ column densities.
The column densities and temperatures we derive from our analysis of the \nicer\ spectra differ from those derived by \cite{2021ApJ...914...47K}. These differences are mostly due to the different methods used in the spectral analysis.  Kashi et al., following \cite{2007ApJ...663..522H}, apparently derived column densities and temperatures by fitting the hard-band portion of the spectrum with a simple absorbed 1-temperature model (see their table 2). Their analysis tends to underestimate the X-ray temperature but overestimate the emission measure and column densities. We use a more realistic model which describes the complex emission seen in the high resolution \chandra\ spectra (as discussed above) and which also includes analysis of lower-energy emission which is most sensitive to column density changes. These temperatures are also consistent with the observed Fe XXVI/Fe XXV ratio $\sim0.1-0.3$, seen in \chandra\ spectra near apastron \citep{2021PhDT.........5E}.

\subsection{X-ray Temperatures}

The maximum X-ray temperature of the shocked gas in a colliding-wind binary should be approximately constant with orbital phase if the pre-shock velocities of the winds are near terminal velocity at every point in the orbit. Figure \ref{fig:eta_car_kT_hot_alone shows the temperatures for the higher-temperature component in our assumed two-temperature spectral model for the CWR X-ray emission.}  
The temperature of this high-temperature component shows significant scatter for most of the orbit, as can be seen in the upper panel, but overall there is not much evidence for significant change in temperature when the stars are well separated.
In the interval 3.5 $<$ $\phi$ $<$ 3.9 the weighted average temperature for this component k$T= 3.84\pm1.05$~keV.  
For comparison,  \cite{2007ApJ...663..522H} derived a temperature of k$T=4.6^{+0.2}_{-0.1}$~keV near $\phi=1.47$.
Over the entire orbit, the weighted average temperature is similar, k$T=  3.74\pm1.06$~keV. 

\begin{figure}[htbp]
  \centering
  \includegraphics[width=\linewidth]{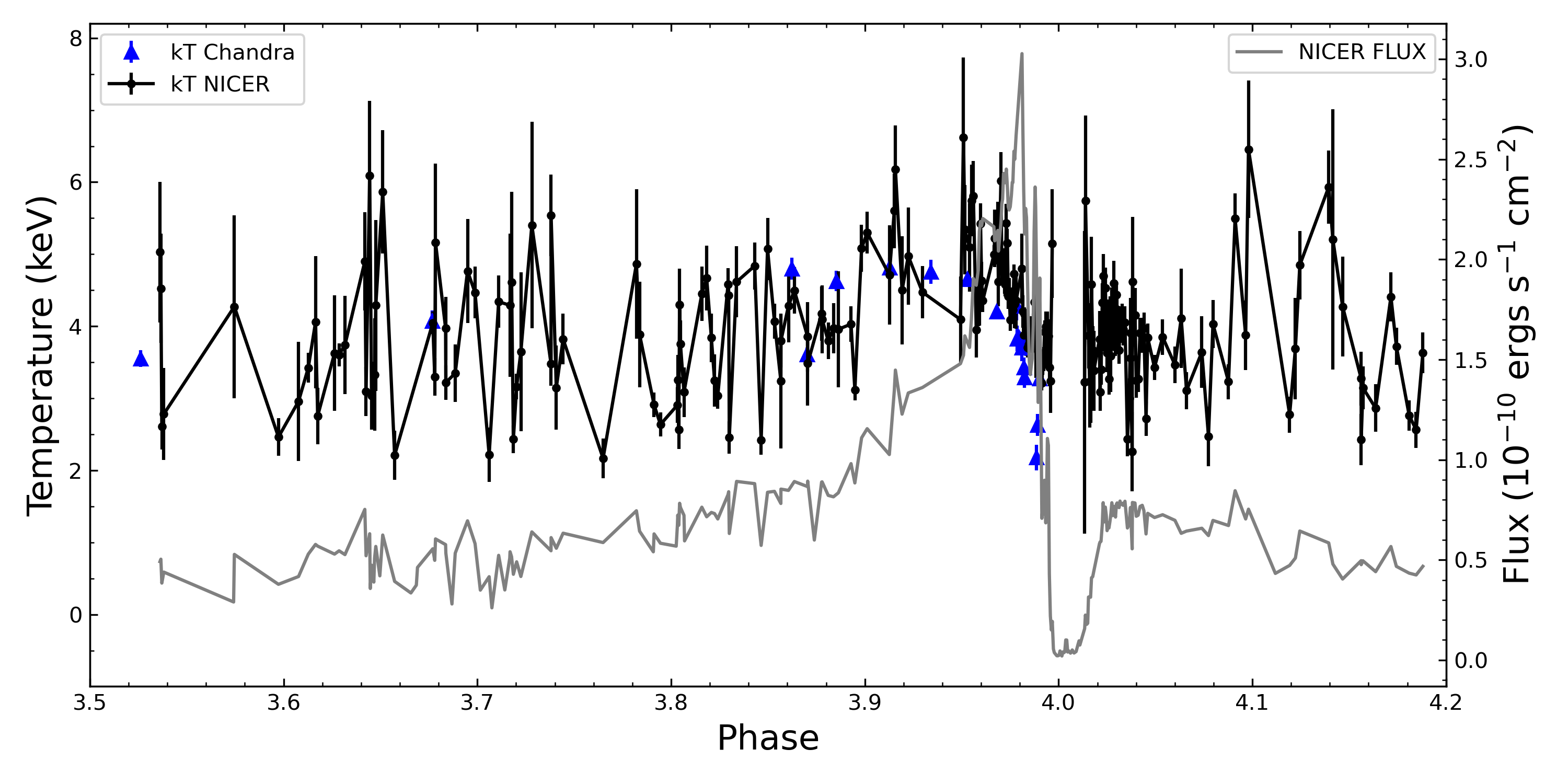}
  \includegraphics[width=\linewidth]{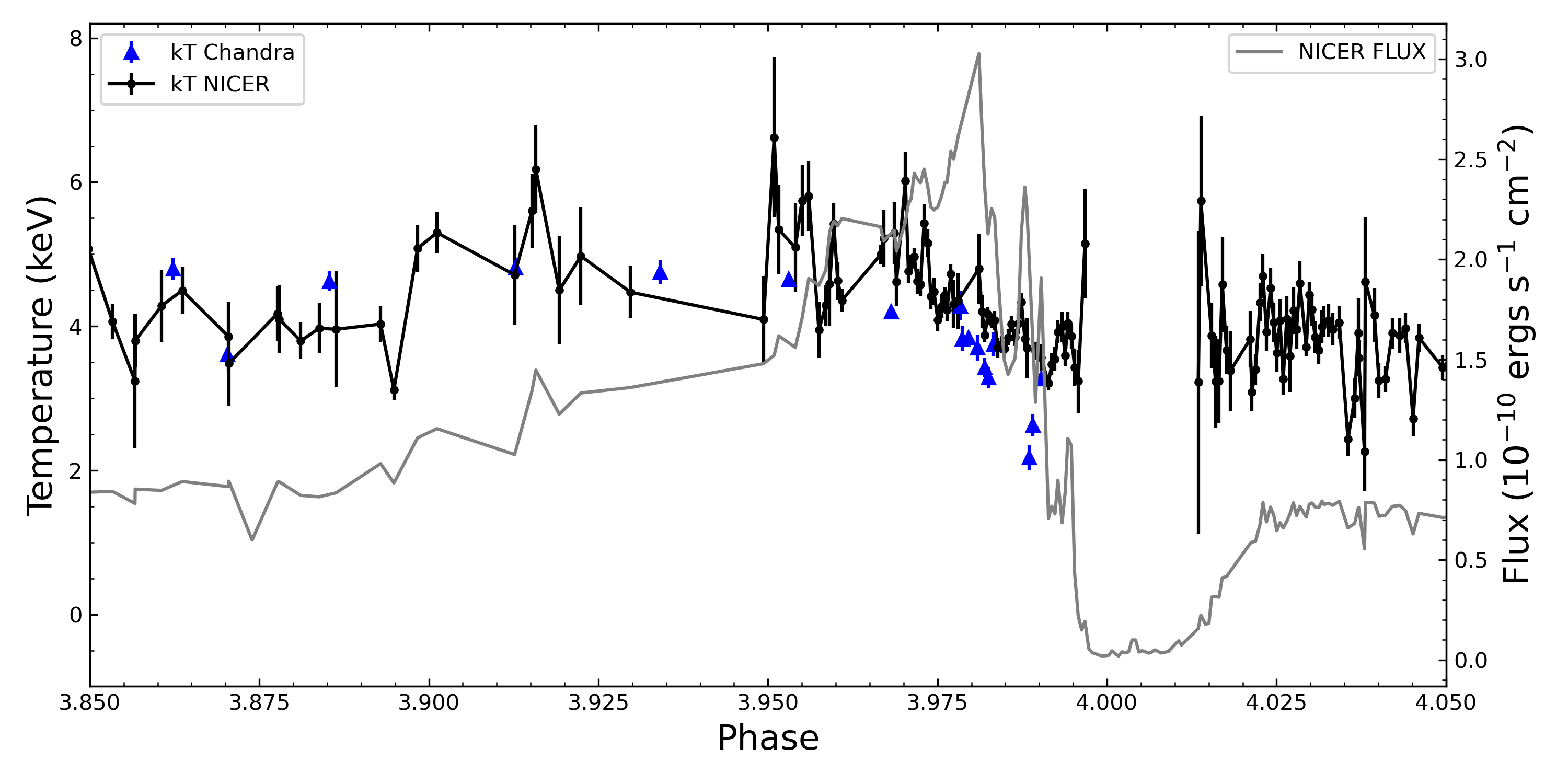}
  \caption{{\it Top:} Temperatures vs phase of the hot component for \ec's \nicer\ observations on this work.
    {\it Bottom:} Same as the top panel near the X-ray minimum ($3.85<\phi< 4.05$). A clear decline in temperature can be seen starting at $\phi\approx 3.97$, a phase when the 2~--~10~keV X-ray flux is still increasing. By $\phi\approx4.02$, the temperature has stabilized around 4.0 keV, with increased variability after $\phi\approx$4.06. Note that the spectrum is too faint during the X-ray minimum ($3.995<\phi<4.015$) for a reliable measure of the X-ray temperatures.
  }
  \label{fig:eta_car_kT_hot_alone}
\end{figure}

The bottom panel in the Figure \ref{fig:eta_car_kT_hot_alone} shows the temperature variation near the X-ray maximum/minimum, in the phase range 3.85 $<$ $\phi$ $<$ 4.05.  Temperature measures are more precise near the X-ray maximum, when the hard emission component is bright.
The temperatures at X-ray minimum are not shown because it is not possible to get reliable measurements from the \nicer\ spectra between 2.0 - 10.0 keV since the source is too faint.
Starting after $\phi>4.10$, the temperature is nearly constant at k$T=3.6\pm0.8$~keV, and again shows more scatter starting at $\phi=4.06$, with an average temperature near 4.0~keV.
As can be seen from Fig.~\ref{fig:eta_car_kT_hot_alone},
there is a significant, apparently monotonic decline in the temperature of this component starting at $\phi=3.96$.
This decline in temperature happens before the plunge of \ec's X-ray emission to minimum, when the stars are moving towards each other as they approach the point of periastron.
The temperature shows a mostly linear decline from about $4.5\pm0.4$~keV (T $=$ 52.2 $\times$ $10^{6}$ K) to approximately $3.0 \pm0.5$~keV (T $=$ 34.8 $\times$ $10^{6}$ K) in only 60 days, cooling at a rate of about 3$\times$ $10^{5}$ K day$^{-1}$.
However, \cite{2007ApJ...663..522H}  measured higher temperatures in the phase range $1.988<\phi<1.99$ from 2 \xmm\ spectra obtained just before the 2003 X-ray minimum, reporting temperatures of k$T=$4.5~keV at 1.988 and  5.4~keV at 1.99, neither of which was consistent with cooling during this interval in 2003.  As a check on the phase dependence of the hot component temperature, we also measured temperatures from \chandra\ HETGS spectra obtained contemporaneously with the \nicer\ observations, using the same spectral model (\citealt{2021PhDT.........5E}, Espinoza Galeas et al., 2021, in prep.)  We found good agreement between the temperatures derived from the fits to the \chandra\ HETGS spectra and the \nicer\ spectra, and in particular that the \chandra\ temperatures showed a similar monotonic decline in the phase interval $3.98-3.99$

\begin{figure}[htbp]
  \centering
  \includegraphics[width=\linewidth]{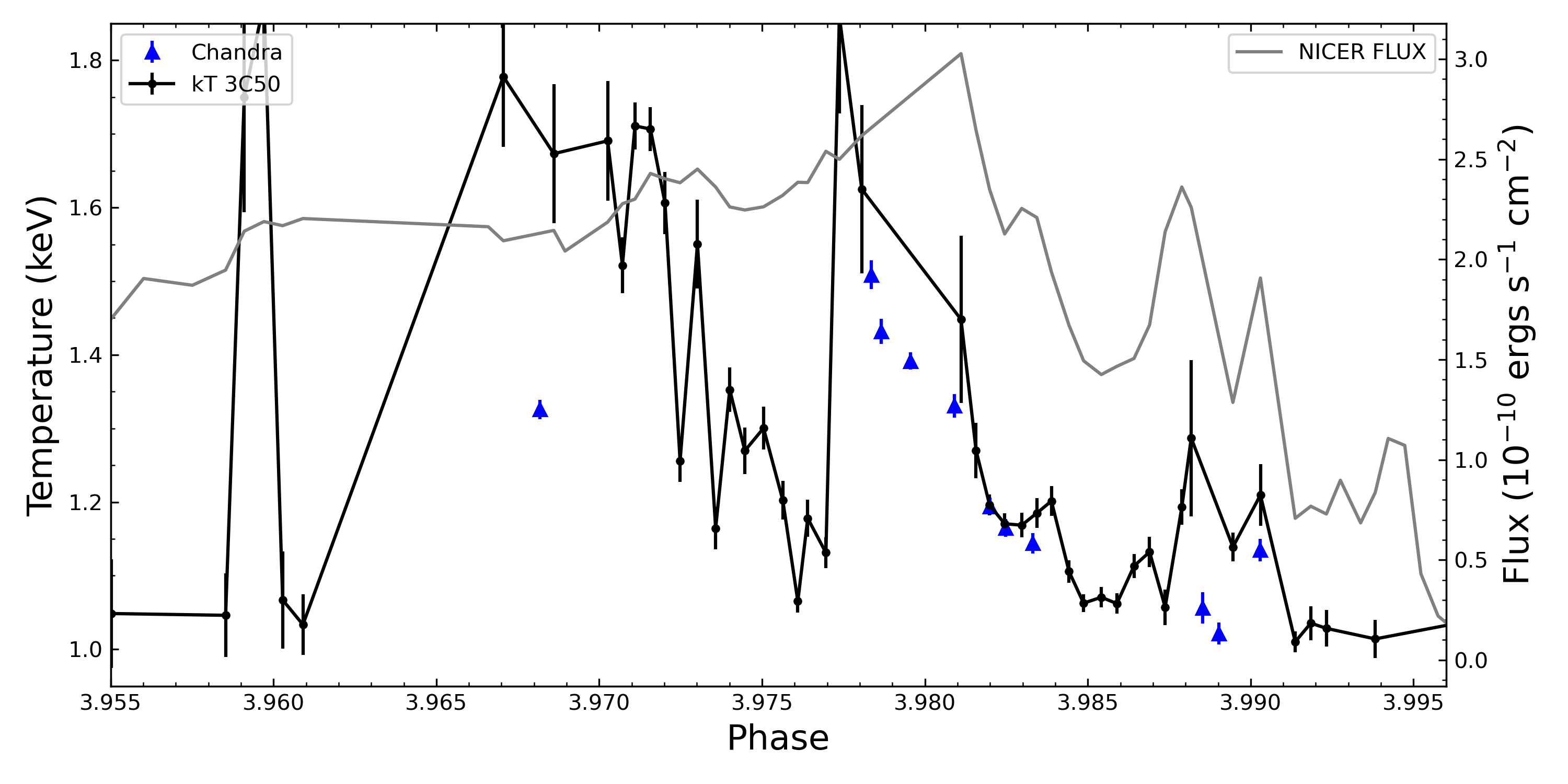}
  \caption{\nicer\ temperatures vs phase for the cool component between phases 3.95 to 4.00.
    The flare region shows an apparent correlation between temperatures and flux, where the peak of the cool component temperature appears to coincide with the flare peaks.  }
  \label{fig:eta_car_kT_cool_alone}
\end{figure}

We also measured the temperature of the cooler component.
The cooler emission originates from gas farther downstream from the shock apex, and is a combination of hotter gas that has cooled as it flows away from the shock apex and lower-temperature shocked gas produced where the wind flows collide indirectly.
The cool component temperatures are often difficult to constrain because of the soft emission from the constant outer debris X-ray emission below 2~keV.  
Several of the fits reach our assumed 1~keV lower constraint.
Temperatures for this cooler component are easier to constrain near the X-ray maximum when the source is bright, but also depend on accurate assessment of  the increasing absorption column.  
In the flare region close to X-ray maximum,  the temperatures of the cool component show interesting behavior, as can be seen in Figure \ref{fig:eta_car_kT_cool_alone}.
At phases between 3.97 and 4.00, the cool component temperatures are somewhat higher than at earlier phases, and show a possible correlation with the flares, where higher flux seems to correspond to higher values of the cool component temperature. 

\section{Discussion}
\label{sec:disc}

\subsection{Changes in Mass-Loss Rates}

The X-ray emission measure and column density provide arguably the most robust measure of the variation in mass loss from the winds of \ec-A and \ec-B.
This is because the 2~--~10~keV thermal emission is dependent on the density of the shocked wind of \ec-B along the shock front, while the absorption of this emission suffered by this emission depends on the density profile of the wind of \ec-A as the X-ray emission travels from the shock front through the wind of \ec-A to the observer.
The excellent agreement in derived column densities between the \nicer\ and \swift\ observations shows that there has been no significant change in the mass-loss rate from \ec-A in the 2014-2020 time interval.
There is also, generally, good agreement between the \nicer\ column densities and most of the column densities from earlier cycles measured by \cite{2007ApJ...663..522H} and \cite{Hamaguchi:2014lr} which suggests, for the most part, a fairly constant density profile through the wind of \ec-A. 
As shown in Figure~\ref{fig:eta_car_nicer_nh}, however, there are some disagreements between the \nicer\ and \swift\ column densities and the earlier measures, in particular in the interval $0.02<\phi<0.05$, in which the \swift\ and \nicer\ column densities are significantly lower than the two measures reported in \cite{2007ApJ...663..522H} in this phase range.
This discrepancy may indicate that a significant change in the wind density profile from \ec-A occurred sometime between 2003 and 2014.
Conclusively deciding the nature of this discrepancy is difficult because of the poor time sampling of the 2003 observations, and stochastic variations in the density profile caused by local clumping in the wind of \ec-A cannot be entirely ruled out (though the discrepancy, about a factor of 5 in derived column at $\phi\approx 0.04$ seems large for a local perturbation which might be produced, for example by a wind clump or other local density enhancement).
At this phase, simulations show that the X-ray emission from the colliding-wind bow shock passes through the distorted wind from \ec-A as the wind is wrapped around the leading wall of the bow shock, and it is plausible that small stochastic changes in this region of the wind could have large effects on the absorbing column.

Overall, the measured X-ray column densities from the 2003-2020 interval agree to better than 40-50\% with previous observations.
Since column density is directly proportional to $\dot{M}$, and since the \nicer\ and \swift\ column densities tend to be lower than the earlier measurements, this agreement constrains  the decline in the mass-loss rate from \ec-A , $\ddot{M_{A}}$, to $\ddot{M_{A}}/\dot{M_{A}}<2.5$ percent per year.

The X-ray flux in the 2~--~10~keV band depends on the emission measure of the shocked wind of \ec-B, which is proportional to the mass-loss rate from the companion, $\dot{M_{B}}$.
Thus comparing fluxes in the 2~--~10~keV band from different orbital cycles provides a measure of the changes in mass-loss rate from the companion, $\ddot{M_{B}}$, over time.
It is best to compare fluxes when the stars are close to apastron, since the shock is stable and cooling is adiabatic.
As shown in Figure~\ref{fig:rxteswiftnicer}, there is good agreement between the \rxte\ Cycle 1 (2003) and Cycle 2 (2009) observations near $\phi \approx 3.8$, while the \rxte\ Cycle 0 and \nicer\ fluxes are about 25\% lower.
This could indicate non-monotonic variability in the mass-loss rate from the companion; also this can indicate some systematic error in calibration between the \rxte\ Cycle 0 and \nicer\ fluxes and the \rxte\ fluxes from the other two cycles.
\cite{2021ApJ...914...47K} suggest that \ec's different X-ray emission recoveries are the result of a decrease in \ec-A's wind. If this were the case we should expect significant differences in column densities for different recoveries. But as shown in Figure \ref{fig:eta_car_nicer_nh}, the column densities in the recoveries observed by \swift\ and \nicer\ are very similar, showing no significant changes in the mass-loss despite the different recovery durations.

\subsection{Shock Temperatures, Wind Velocities, and the Radius of \ec-B}

As seen in Figure~\ref{fig:eta_car_kT_hot_alone}, the hotter component shows a fairly constant X-ray temperature of k$T= 3.84\pm1.05$~keV up to $\phi\approx 3.96$, after which there is a rather linear decline up to $\phi=3.99$ (at which point the spectra become too faint for analysis).
The hottest plasma in the colliding-wind shock  originates near the apex of the shock cone, where the winds collide nearly head on, and as the stars approach periastron passage,  the apex of the colliding-wind shock moves closer to the companion star.
If the apex enters the region of the companion's wind where the wind is still accelerating and has not yet reached terminal velocity, this would cause a drop in the temperature of the hot shocked gas near the apex.
The drop in temperature of $\approx$2/3 by $\phi=3.99$ compared to the temperature at $\phi=3.96$ when the decline apparently starts, suggests a drop in the pre-shock velocity of the fast wind of \ec-B. 
Since $T\propto V^{2}$, this corresponds to a decrease in the pre-shock velocity of the wind of \ec-B of $(V_{1} - V_{2})/V_{1} \approx$ 2.5\% where $V_{1}$ is the wind velocity at $\phi=3.96$ and $V_{2}$ is the wind velocity $\phi=3.99$.  

 \begin{figure}[htbp]
  \centering
  \includegraphics[width=0.6\linewidth]{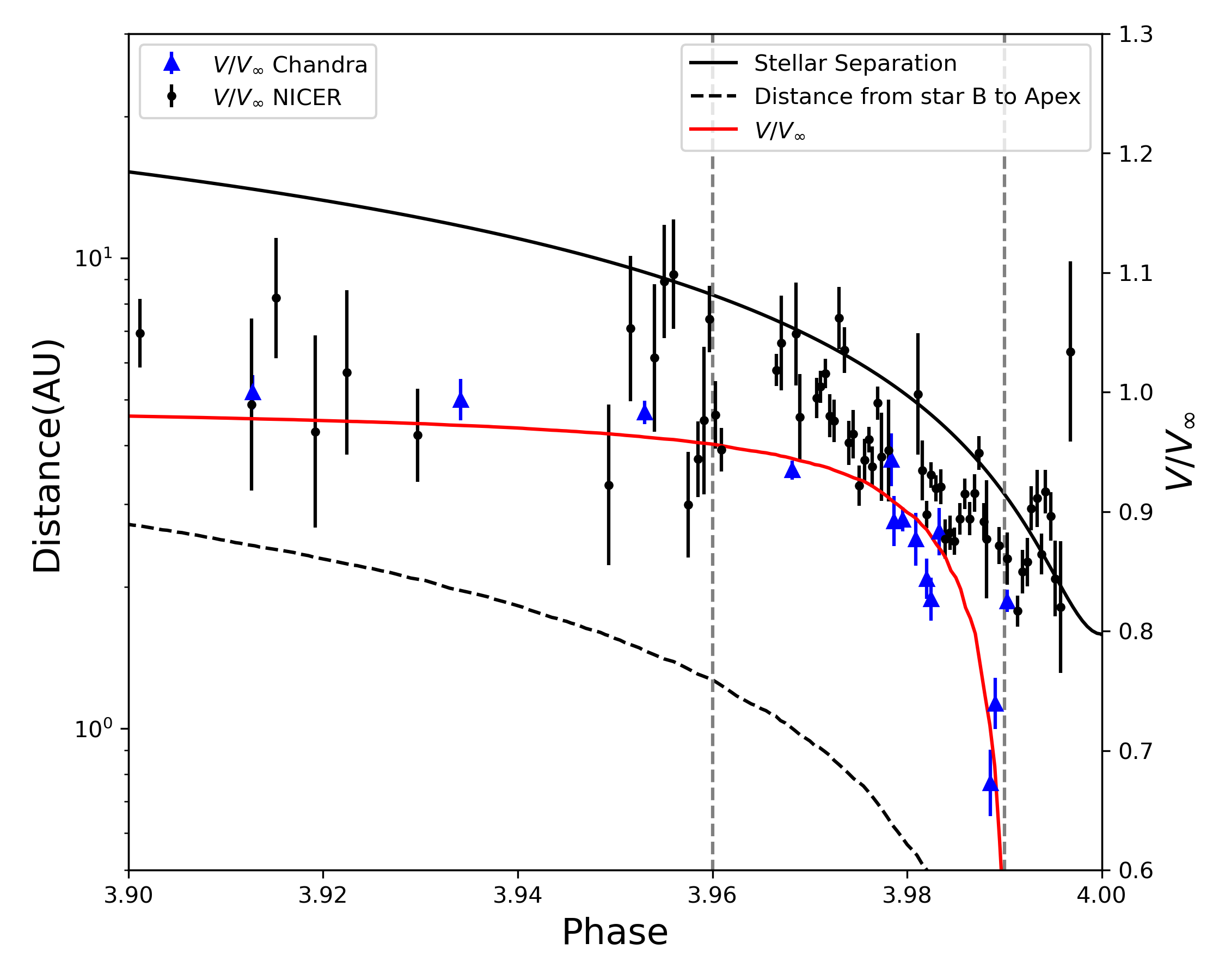}
  \caption{The Figure shows the expected variation in stellar separation (solid black line) with the $v/v_{\infty}$ of \ec~B's wind at the location of the apex (red line), derived using the system and stellar parameters given in Table \ref{tab:sysparams}, and a radius of \ec-B = 30~\rs.
  The gray vertical dashed lines indicate the observed phase interval of the decline in the X-ray temperature of the hot component.
}
\label{fig:dvdt}
\end{figure}

Figure~\ref{fig:dvdt} shows the expected variation in separation of the two stars and the distance from \ec-B to the stagnation point of  the shock, derived by balancing the pressure of the wind from \ec-A and \ec-B between the two stars.
Figure~\ref{fig:dvdt} also shows the derived change in wind velocity assuming a standard wind velocity law $V(r) = V_{\infty, B}(1-R_{B}/r)^{\beta}$, where $V(r)$ is the velocity of \ec-B's wind at a distance $r$ from the star, $V_{\infty, B}$ is the terminal velocity of \ec-B's wind, and $\beta$ the wind acceleration constant (usually close to 1 for most radiatively-driven winds).
Note that very near periastron, the apex moves very close to the companion and the velocity calculation breaks down.
The dashed vertical gray lines indicate the phase interval of the observed cooling of the hotter component, while the red line shows the change in wind velocity implied by the temperature decreases, assuming $T\propto V^{2}$.
The intersection of the dashed vertical gray and red line show the derived velocities at the start and end of the observed cooling interval.
We find that the derived change in wind velocity is fairly well matched by the calculated variation in \ec-B's pre-shock wind velocity at the distance of the shock cone apex over this range of orbital phase, if the radius of \ec-B is $\approx 30$~\rs.  
Ionization modeling by \cite{2005ApJ...624..973V} yielded values of $T_{eff,B}= 37,200$~K and $\log L_{B}/L_{\odot} \approx 5.97$, implying an effective radii of 23.6~\rs, slightly smaller than the radius which reasonably describes the X-ray cooling if $\beta=1$ shown in Figure~\ref{fig:dvdt}; but, given the uncertainties and assumptions, these numbers are in good agreement.
On the other hand, photoionization modeling by \cite{Mehner:2010rm} suggests a somewhat hotter but fainter companion star, $T_{eff,B}= 39,000$~K and $\log L_{B}/L_{\odot} \approx 5.60$, yielding an effective radius of only 14.0~\rs.
We can reasonably describe the observed X-ray cooling curve with these radii, but both require higher values of $\beta$ (indicating a more slowly accelerating wind), $\beta\approx 1.5$ and $\beta\approx 3.0$ for the Verner et al. and Mehner et al. radii, respectively.

\subsection{The Deep/Shallow Minimum Transition}

The \nicer\ campaign has provided the best measure obtained so far of the variations in X-ray  flux and X-ray spectral properties around the X-ray minimum.
The \nicer\ HR in Figure \ref{fig:eta_car_nicer_hr} shows the transition between the ``deep'' and ``shallow'' states of the X-ray minimum first identified by \cite{Hamaguchi:2014lr} from a set of only five \chandra\ ACIS spectra.
Figure \ref{fig:eta_car_nicer_hr} also show that the deep minimum interval lasts for an interval of 18 days, starting at $\phi=3.995$ and ending at $\phi=4.004$, with the transition to the shallow minimum.
The shallow interval ends at $\phi=4.013$, a duration of 18 days while the shock gradually strengthens and the absorption dissipates as the companion moves around and away from \ec-A.
This is consistent with the interpretation offered in  \cite{Hamaguchi:2014lr}, who suggested that the deep minimum is produced by occultation of the X-ray emitting gas by the optically-thick inner wind of \ec-A, which completely hides the X-ray emission from the colliding-wind shock, plus a decline in X-ray emissivity near periastron passage, followed by a gradual strengthening of the shock and a re-heating to X-ray emitting temperatures.  

\subsection{The Column Density Variation Through \ec's Orbital Motion}
The column density from the hot component of \ec\ X-ray emission shows variability through all the \nicer\ monitoring program. 
Those changes go up to one order of magnitude in column density (Figure \ref{fig:eta_car_nicer_nh}), with intervals no longer than $\sim0.1$ in phase.
The variations indicate changes in the amount of material in front of the apex of the CWR in our line of sight.
The occurrence and duration of the $N_{H}$ variations before $\phi=3.95$ do not show a particular correlation with flux or phase.
Stochastic changes in \ec~A's wind by clumps could explain the variability observed.
The lack of more frequent observations before $\phi=3.95$ makes difficult a deeper analysis of those variations in column density. 
% constrain in the time that these increases last.

After $\phi=3.94$ more frequent \nicer\ observations were allocated.
Between phases 3.96 and 3.97 the $N_{H}$ shows a particular stability in the measurements. 
The stability of so many subsequent measurements was not observed before.
The measurements of $N_{H}$ between $3.97<\phi<3.98$ makes clear the increase of one order of magnitude in column density after phase 3.98.
After $\phi=3.98$ the column density keeps almost constant for the next $\Delta\phi\approx0.1$.
The step of one order of magnitude at $\phi=3.97$ can be also due a clump in \ec~A wind. 
Another interesting explanation for the step change in column density is a transition between the wind from \ec~B and \ec~A separated by the walls of the cone formed by the CWR.

\subsection{Variation in the Duration of the X-ray Minimum}
As first noted by \cite{2020ATel13636....1E}, \nicer\ showed that the duration of the X-ray minimum in 2020 was the shortest seen in any of the four orbital cycles monitored so far.  The first two minima seen by \rxte\ in 1997 and 2003 had nearly identical X-ray minimum durations, while the \rxte\ monitoring of the 2009 minimum had a significantly shorter duration.  
Any hypothesis needs to explain the apparently stochastic behavior observed in the starting time of the X-ray recovery observed in Figure \ref{fig:rxteswiftnicer}.
\cite{2010ApJ...725.1528C} suggested that the different recoveries are due to a decline in \ec's wind momentum.
The decline can be caused by a drop in mass-loss or wind terminal velocity, or some combination.
If so, we might expect to see differences in column densities when comparing X-ray spectra obtained during recovery in different orbital cycles.
Figure \ref{fig:eta_car_nicer_nh} compares the column densities from cycle 3 (from \swift) and cycle 4 (\nicer).
Although the duration of the cycle 3 minimum was substantially longer than the cycle 4 recovery (See Figure \ref{fig:rxteswiftnicer}), after $\phi\approx4.020$ the column densities from \swift\ and \nicer\ are very similar.
This suggests that the differences in the duration of the cycle 3 and cycle 4 minima was not produced by a significant change in the wind momentum of the primary.
\cite{2021ApJ...914...47K} and \cite{2009ApJ_701_L39} propose that accretion at periastron weakens the mass-loss rate of \ec-A every cycle, implying that any new X-ray emission recovery should be shorter than the previous one. But as discussed above it is not likely that \ec-A's wind has weakened significantly from 2015 to 2020 based in the column densities. Also, the 2009 minima was shorter than the 2015 minima as shown in Figure \ref{fig:rxteswiftnicer}, indicating that there is no systematic weakening of \ec-A's wind with time.

The comparison of the \nicer\ and \swift\ observations around X-ray minimum suggests that another mechanism produces the variation in duration of the X-ray recoveries.
Winds from luminous hot stars are prone to line-deshadowing instabilities (LDIs) that can randomly create localized variations in wind density \citep{1988ApJ...335..914O}.
Figure 4 in \cite{1988ApJ...335..914O} shows how those instabilities can increase the velocity of the wind from $\sim$500~km/s to $\sim$1250~km/s, especially in the dense inner parts of the wind.
Such local density enhancements (clumps) interacting with the colliding-wind bow shock can move the stagnation point of the CWR closer to \ec~B, reducing the pre-shock wind speed and reducing the hot X-ray flux from the bow shock.
In this scenario, we can still have regions where the shock of the winds can generate high energy X-rays but distributed over a broader region, that together with a high column density, produces the shallow minimum.
The shallow minimum duration will be dependent on the size of the clump and the time when the clump forms.
Once the clump passes the shock, \ec~B's wind can accelerate enough to re-form the hot shocked gas in the CWR between the stars, starting the recovery of the hard X-ray flux.

We suggest that the sudden, step-like increase in column density before the X-ray minimum (See Figure \ref{fig:eta_car_nicer_nh}) indicates the formation of a dense clump produced by the LDIs in the inner wind of \ec~A.
This step is observed at a later phase in cycle 3 than in cycle 4, suggesting a connection between the clump formation time and the time of recovery. 
Clumps that form later produce a longer X-ray minimum, while clumps which form earlier in orbital phase pass beyond the bow shock earlier, producing an earlier X-ray recovery and shorter X-ray minimum.

%The different recoveries is a puzzle that probably requires an explanation with a combination of several effects occurring at periastron passage.
%The presence of flares close to periastron passage and the step function in NH before X-ray minimum indicate changes in the hydrodynamic conditions that can affect the recovery of the CWR.
%\cite{2009ApJ...707..693M} include in their discussion about flares, the possibility of clumps in \ec's wind produced by radiative wind instabilities or subsurface convection. 

\subsection{The Nature of Rapid X-ray Variability}

As discussed in Section \ref{sec:Flaring}, \nicer\ observed similar rapid X-ray variability, or ``flares'', which become evident near X-ray maximum and are even seen during the  decline to minimum.
The comparison of the \nicer\ 2~--~10~keV X-ray fluxes with those from \rxte\ Cycle 1 and Cycle 2 as a function of time shows no strong correlation.
%that there are four \nicer\ flares which occur at very similar phases as the flares seen in either Cycle 1 or Cycle 2.
%Given the different time sampling of the different cycles this coincidence is rather remarkable.
\cite{2009ApJ...707..693M} concluded that this flaring activity was produced by large, homologously-expanding localized overdensity regions, or clumps, in the wind of \ec-A.

The apparent coincidence of peaks in the temperature of the cool component with the peaks of some X-ray flares seen in Figure~\ref{fig:eta_car_kT_cool_alone} can be interpreted in terms of the ``clump'' model of \cite{2009ApJ...707..693M}.
In this model, the pressure of a clump on the wall of the bow shock downstream from the apex could in principle decrease the opening angle of the bow shock, causing the winds to collide more directly, increasing the temperature of the downstream shocked wind while increasing the density of that portion of the shock producing an increase in X-ray flux from the shocked wind of the companion star.  

\section{Conclusions and Future Work}
\label{sec:conc}
\nicer\ provides time-resolved measures of the 0.4-10 keV \ec's X-ray spectrum from $\phi\sim$3.5 and ongoing.
Flux measured by \nicer\ follows the $1/D$ behavior for most of the orbit, similar to \rxte\ and \swift\ lightcurves.
\nicer\ observed that the plunge of the X-ray maximum started at $\phi\sim$3.98, similar to the previous cycles indicating that it is strongly correlated with orbital phase. 
Inspection of \ec's \nicer\ spectrum does not show evidence of a decrease in X-rays due to absorption, suggesting that the decrease in flux is due to a decrease in temperatures at the CWR.
This is the first time we have evidence of temperature decrease in the CWR of \ec.
The temperature of the hot component declines on approach to periastron passage, indicating that the shock apex is moving into the acceleration zone of the wind of \ec~B.

Measurements of the HRs with \nicer\ constrain the Deep minimum to 3.995$<\phi<$4.004 (18 days) and Shallow minimum from 4.004$<\phi<$4.013.
We have observed the shortest X-ray recovery: the flux starts to increase at $\phi=4.009$ which is $\sim$7 days earlier than the low limit estimated by \cite{2010ApJ...725.1528C} in the 2009 periastron passage.
The apastron fluxes do not change $>$5\%, indicating a change no more than 0.25\% in $\dot{M}$ from \ec~A or B.
The absorption measured by \nicer\ shows an agreement up to 90\% after $\phi=4.01$, giving us another indicator that $\dot{M}$ from \ec~A or \ec~B has not changed significantly in the last two cycles.

The changes in the soft-band emission from the OE seen for the first time by \nicer\ may be caused by the expansion of the ejecta. If so, simple analysis indicates that the X-ray luminosity near the time of the Great Eruption was about $\sim10^{41}$ \lumcgs. This is the first estimate of the X-ray luminosity of the Great Eruption and suggests that the X-ray luminosity at that time was comparable to the total luminosity at longer wavelengths.

\acknowledgements
D. Espinoza Galeas gratefully acknowledges support from NASA grants  \#80NSSC19K1451 and  \#80NSSC21K0092, and SAO grant \#GO9-20015A thru NASA.
This work was conducted as part of doctoral research conducted at The Catholic University of America.
M. F. Corcoran and K. Hamaguchi are supported under the CRESST-II cooperative agreement  \#80GSFC17M0002 with the NASA/Goddard Space Flight Center.
C. M. P. Russell  was supported by SAO grant \#GO0-21006A  through NASA; this support is gratefully acknowledged.  This research has made use of data and software provided by the High Energy Astrophysics Science Archive Research Center (HEASARC), which is a service of the Astrophysics Science Division at NASA/GSFC.  This research has made use of NASA’s Astrophysics Data System. This research made use of Astropy,\footnote{http://www.astropy.org} a community-developed core Python package for Astronomy \citep{astropy:2013, astropy:2018}.

\software{XSPEC (Arnaud 1996),
astropy (The Astropy Collaboration 2013, 2018)}

\bibliography{ecnicer.bib}
\bibliographystyle{aasjournal}

\appendix

\startlongtable
\begin{deluxetable*}{ccccrcccccc}
\tabletypesize{\scriptsize}
\tablecaption{Log of \textit{NICER} Observations and Rates \label{tab:nicer-rates}}
\tablehead{\\ \multicolumn{5}{c}{} & \multicolumn{2}{c}{$2.0-10.0$~keV}& \multicolumn{2}{c}{$2.0-3.0$~keV}& \multicolumn{2}{c}{$5.0-7.0$~keV}\\ \hline \colhead{OBSID} & \colhead{MJD Start} & \colhead{MJD End} & \colhead{Phase} & \colhead{EXPOSURE} & \colhead{Total} & \colhead{Net} & \colhead{Total} & \colhead{Net} & \colhead{Total} & \colhead{Net}\\ \colhead{ } & \colhead{ } & \colhead{ } & \colhead{ } & \colhead{$\mathrm{s}$} & \colhead{cts/s} & \colhead{ } & \colhead{cts/s} & \colhead{ } & \colhead{cts/s} & \colhead{ } }\startdata
1110010101 & 57954.1245 & 57954.9641 & 3.5343 & 299.7 & 6.09 & $5.55\pm0.14$ & 2.66 & $2.54\pm0.09$ & 0.71 & $0.60\pm0.05$ \\
1110010102 & 57955.0252 & 57955.9950 & 3.5347 & 677.0 & 6.69 & $5.83\pm0.10$ & 2.83 & $2.67\pm0.06$ & 0.83 & $0.64\pm0.04$ \\
1110010103 & 57956.1829 & 57956.8311 & 3.5352 & 636.5 & 6.59 & $4.94\pm0.10$ & 2.82 & $2.54\pm0.07$ & 0.82 & $0.44\pm0.04$ \\
1110010105 & 57958.4340 & 57958.6970 & 3.5362 & 392.9 & 6.51 & $5.39\pm0.13$ & 2.70 & $2.48\pm0.08$ & 0.81 & $0.55\pm0.05$ \\
1110010106 & 58031.0912 & 58031.8693 & 3.5723 & 297.1 & 7.07 & $3.47\pm0.15$ & 3.15 & $2.63\pm0.10$ & 0.85 & $<0.05$ \\
1110010107 & 58032.2539 & 58032.2613 & 3.5727 & 199.5 & 9.73 & $6.33\pm0.22$ & 3.45 & $2.93\pm0.13$ & 1.44 & $0.62\pm0.09$ \\
1110010108 & 58078.2795 & 58078.3071 & 3.5954 & 694.8 & 6.79 & $4.92\pm0.10$ & 2.87 & $2.56\pm0.06$ & 0.85 & $0.42\pm0.04$ \\
1110010109 & 58099.2565 & 58099.3248 & 3.6058 & 163.3 & 6.74 & $5.40\pm0.20$ & 2.99 & $2.73\pm0.14$ & 0.80 & $0.49\pm0.07$ \\
\enddata
\end{deluxetable*}

\label{tab:nicer-rates}

\newpage

%PUT nicerSpectab.tex here

\startlongtable
\begin{deluxetable*}{cccccccccc}
\centerwidetable
\tabletypesize{\scriptsize}
\tablecaption{\textit{NICER} Spectral Modeling Parameters \label{tab:nicer-spectab}}
\tablehead{\colhead{OBSID} & \colhead{Phase} & \colhead{Flux} & \colhead{NH$_{3}$} & \colhead{kT$_{4}$} & \colhead{EM$_{4}$} & \colhead{NH$_{5}$} & \colhead{kT$_{6}$} & \colhead{EM$_{6}$} & \colhead{Flux$_{FeK}$}\\ \colhead{ } & \colhead{ } & \colhead{10$^{-11}$ ergs cm$^{-2}$ s$^{-1}$} & \colhead{10$^{22}$ cm$^{-2}$} & \colhead{$\mathrm{keV}$} & \colhead{10$^{58}$ cm$^{-3}$} & \colhead{10$^{22}$ cm$^{-2}$} & \colhead{$\mathrm{keV}$} & \colhead{10$^{58}$ cm$^{-3}$} & \colhead{10$^{-5}$ photons cm$^{-2}$ s$^{-1}$}}
\startdata
1110010101 & 3.53 & 4.96 & $3.24\pm0.25$ & $1.01\pm0.11$ & $0.74\pm0.18$ & $3.43\pm0.89$ & $5.91\pm1.18$ & $0.30\pm0.04$ & $6.46\pm4.40$ \\
1110010102 & 3.53 & 5.03 & $3.25\pm0.14$ & $1.32\pm0.09$ & $0.97\pm0.10$ & $10.77\pm2.45$ & $4.40\pm0.83$ & $0.40\pm0.08$ & $5.83\pm3.08$ \\
1110010103 & 3.54 & 3.84 & $3.37\pm0.44$ & $<1$ & $0.50\pm0.18$ & $3.43\pm0.46$ & $2.64\pm0.42$ & $0.50\pm0.14$ & $1.58\pm2.87$ \\
1110010105 & 3.54 & 4.35 & $3.59\pm0.17$ & $1.14\pm0.09$ & $1.21\pm0.15$ & $14.52\pm4.55$ & $2.43\pm0.71$ & $0.89\pm0.53$ & $6.38\pm3.94$ \\
1110010106 & 3.57 & 2.65 & $3.73\pm0.37$ & $1.29\pm0.15$ & $1.31\pm0.43$ &  &  &  &  \\
1110010107 & 3.57 & 5.15 & $3.00\pm1.37$ & $1.00\pm0.46$ & $0.29\pm0.19$ & $3.00\pm1.24$ & $3.52\pm0.90$ & $0.49\pm0.13$ & $9.54\pm8.21$ \\
1110010108 & 3.60 & 3.57 & $3.00\pm4.25$ & $1.30\pm0.59$ & $0.12\pm0.12$ & $3.00\pm0.71$ & $2.13\pm0.15$ & $0.66\pm0.10$ & $2.09\pm2.98$ \\
1110010109 & 3.61 & 3.93 & $3.00\pm2.36$ & $1.51\pm0.39$ & $0.90\pm0.60$ & $3.44\pm1.23$ & $2.47\pm0.51$ & $0.54\pm0.18$ & $4.98\pm5.77$ \\
\enddata
\end{deluxetable*}

\end{document}